\documentclass[12pt]{article}
\usepackage[top=1in, bottom=1in, left=1.in, right=1.in]{geometry}
\usepackage[english]{babel}
\usepackage{atbegshi,cite}
\usepackage{amsmath,amssymb,amsbsy,amstext, amsthm, simplewick}
\usepackage{hyperref}
\usepackage{graphicx}
\usepackage{amsfonts}
\usepackage[small]{caption}
\usepackage{upgreek}
\usepackage[titletoc]{appendix}
\usepackage{setspace}
\usepackage[dvipsnames]{xcolor}
\usepackage{slashed}

\newcommand{\al}{\alpha}
\newcommand{\bt}{\beta}
\newcommand{\g}{\gamma}

\newcommand{\dt}{\delta}

\newcommand{\simu}{\sigma^{\mu\nu}}

\newcommand{\vL}{\ensuremath{\mathcal{L}}}
\newcommand{\vp}{\varphi}
\newcommand{\sq}{^{2}}

\newcommand{\dslash}[1]{#1 \llap{/\kern-0.5pt}}
\newcommand{\Dslash}[1]{#1 \llap{/\kern+1.5pt}}
\newcommand{\DDslash}[1]{#1 \llap{/\kern+2.3pt}}
\newcommand{\dslashh}[1]{#1 \llap{/\kern+1pt}}
\newcommand{\abs}[1]{|#1|}

\newcommand{\op}{\mathcal O}

\newcommand{\Ex}[1]{\cdot 10^{#1}}
\newcommand{\bea}{\begin{eqnarray}}
\newcommand{\eea}{\end{eqnarray}}
\newcommand{\bma}{\begin{pmatrix}}
\newcommand{\ema}{\end{pmatrix}}
\newcommand{\nn}{\nonumber}

\newcommand{\Tr}{\mathrm{Tr}}
\renewcommand{\Im}{\mathrm{Im}}

\theoremstyle{plain}
\theoremstyle{plain} 
\theoremstyle{plain} 
\theoremstyle{plain}
\theoremstyle{plain}
\theoremstyle{plain}

\renewcommand{\title}[1]{{\Large\bf\flushleft{#1}}\vspace*{3ex}\\}
\renewcommand{\author}[2]{{\noindent\hspace*{2.5em}\large#1}
                     \footnote{Electronic mail: $\mathtt{#2}$}\\}

\begin{document}
\begin{titlepage}
\begin{flushright}
LA-UR-21-26248
\end{flushright}

\vskip 2.2cm

\begin{center}

{\large \bf Uncovering an Axion Mechanism with the EDM Portfolio }

\vskip 1.4cm

{{Jordy de Vries}$^{(a,b)}$, {Patrick Draper}$^{(c)}$, {Kaori Fuyuto}$^{(d)}$, {Jonathan Kozaczuk}$^{(e)}$, and  {Benjamin Lillard}$^{(c)}$}
\vskip 1cm
{$^{(a)}$ {\it Institute for Theoretical Physics Amsterdam and Delta Institute for Theoretical Physics, University of Amsterdam, Science Park 904, 1098 XH Amsterdam, The
Netherlands}}\\
{$^{(b)}$ {\it Nikhef, Theory Group, Science Park 105, 1098 XG, Amsterdam, The Netherlands}}\\
{$^{(c)}$ {\it Department of Physics, University of Illinois, Urbana, IL 61801}}\\
{$^{(d)}$ {\it  Theoretical Division, Los Alamos National Laboratory, Los Alamos, NM 87545, USA}}\\
{$^{(e)}$ {\it  Department of Physics, University of California, San Diego, CA 92093, USA}}\vspace{0.3cm}
%
\vskip 1.5cm

\begin{abstract}
Effective field theory arguments suggest that if BSM sectors contain new sources of CP-violation that couple to QCD, these sources will renormalize the $\theta$ term and frustrate ultraviolet solutions to the strong CP problem. Simultaneously, they will generate distinctive patterns of low-energy electric dipole moments in hadronic, nuclear, atomic, and molecular systems. Observing such patterns thus provides evidence that strong CP is solved by an infrared relaxation mechanism. We illustrate the renormalization of $\theta$ and the collections of EDMs generated in a several models of BSM physics, confirming effective field theory expectations, and demonstrate that measurements of ratios of electric dipole moments at planned experiments can provide valuable input on the resolution of the strong CP problem.
\end{abstract}

\end{center}

\vskip 1.0 cm

\end{titlepage}
\setcounter{footnote}{0} 
\setcounter{page}{1}
\setcounter{section}{0} \setcounter{subsection}{0}
\setcounter{subsubsection}{0}
\setcounter{figure}{0}


\section{Introduction}

The universe violates parity (P) and charge-parity (CP) symmetries. In the standard model (SM), the weak interactions break P and CP, while beyond the standard model,  new sources of CP violation are required to generate the baryon asymmetry. For these reasons, it is a surprise that the strong interactions appear to be P/CP symmetric. Even at the renormalizable level, the strong interactions contain a P/CP-violating phase $\bar\theta$. The limit on the neutron electric dipole moment (EDM), however, presently places a bound $\bar\theta < 1.2 \times 10^{-10}$~\cite{Abel:2020gbr,Dragos:2019oxn}. If P and CP are not symmetries of the universe, why, then, does the strong force appear to conserve these quantum numbers to fantastic precision? Under renormalization, we might expect an O(1) $\bar\theta$ to be generated, even if for some reason it vanishes as an ultraviolet (UV) boundary condition.

It is a curious fact that this na\"ive renormalization group expectation is not reflected by the SM alone, even if it appears to be true in generic extensions of the SM. Radiative corrections to $\bar\theta$ in the SM start at high loop order~\cite{ellisgaillard} and are estimated to be well below the current experimental bound. This surprising property is the starting point for UV solutions to the strong CP problem, including Nelson-Barr models based on spontaneous CP violation, models based on spontaneous parity violation, and others~\cite{Nelson:1983zb,Barr:1984qx,bbp,mohapatrasenjanovic,Beg:1978mt,Georgi:1978xz,Babu:1989rb,barrsenjanovic}. If the UV Lagrangian preserves P, CP, or suitable generalizations of them, then $\bar\theta=0$ is a natural UV boundary condition. These symmetries must be broken spontaneously, and the trick is to sequester the order parameters sufficiently from QCD so that $\bar\theta$ is not regenerated when the spontaneous symmetry breaking sector is integrated out. This is a subtle model-building problem, because the fact that $\bar\theta$ is negligibly renormalized in the SM appears to be quite special, due to the limited flavor structures of the SM, and is not a property shared by most extensions of the SM~\cite{Dine:1993qm,hillerschmaltz,Dine:2015jga,Albaid:2015axa,Draper:2016fsr,Draper:2018tmh,deVries:2018mgf}. Minimal left-right models, for example, preserve parity in the UV, so that $\bar\theta=0$ is a natural boundary condition, but it is regenerated after electroweak symmetry breaking by new phases in the Higgs sector. However, various classes of  UV solutions are known that might achieve the required sequestration, in which case the special radiative structure of the SM takes over and protects $\bar\theta$ into the infrared (IR).

Generically, one does not expect to observe any new sources of hadronic CP violation if strong CP is addressed by a UV mechanism. The sequestering required to protect $\bar\theta$ from spontaneous P/CP breaking to several loop order also typically prevents any other sources of hadronic CP violation -- various dimension-six operators, in the language of the SM effective field theory (SMEFT) -- from being generated at an observable level. This  can be verified in explicit models for UV solutions. For example, the two-loop corrections to $\bar\theta$ in models of softly broken generalized parity typically dominate the pattern of EDMs~\cite{MBinprep}. More generally, the argument can be sharpened examining the divergence structure of SMEFT~\cite{deVries:2018mgf}. dimension-six CPV operators coupling to quarks or gluons give rise to one-loop quadratic divergences in $\bar\theta$, reflecting strong sensitivity of $\bar\theta$ to the CPV physics at the cutoff.

Contrapositively, UV models proposed to address strong CP can be constrained by experimental evidence for new (non-$\bar\theta$) sources of hadronic CP violation. The argument is one of naturalness: even if $\bar\theta$ is small in the UV, if  other sources of hadronic CP violation are present at an observable level, it is difficult to understand from an effective field theory point of view why $\bar\theta$ should be small in the IR.  Thus, if new sources of strong-sector CP violation are observed,  axion solutions to the strong CP problem -- which largely relax the IR value of $\bar\theta$ regardless of its renormalization at intermediate scales and the presence of higher dimension operators -- become essentially the only game in town.

This EFT argument gives low energy EDM experiments a unique and interesting window into the strong CP problem. These experiments are the most sensitive probes of new sources of flavor-diagonal CPV beyond the SM. A complication, however, is that a small bare $\bar\theta$ itself generates a pattern of hadronic EDMs. Therefore, correlated measurements are required to distinguish a pure $\bar\theta$ scenario from one with distinct new CPV sources. Given a sufficient suite of hadronic EDM measurements, if nonzero EDMs are observed, it is possible to disentangle the various contributions from a small nonzero $\bar\theta$ and other CPV sources. If the latter are  present, the argument above can be applied.

In this work we illustrate how this EFT argument operates by studying a few  concrete UV extensions of the SM. These models introduce new CP-violating interactions for quarks and gluons somewhere above the weak scale. We show that these interactions renormalize $\bar\theta$ at tree level or one loop, necessitating an IR solution to the strong CP problem. Furthermore, we show that the models  generate distinctive correlated patterns of EDMs distinguishable from the pattern generated by $\bar\theta$ alone. 

This paper is organized as follows. In Sec.~\ref{theta-renorm} we discuss the renormalization of $\bar \theta$ in the SM and beyond. We recap how $\bar \theta$ is corrected in SMEFT and introduce three BSM models with additional sources of CP violation that explicitly verify the EFT expectations. In Sec.~\ref{EDMpheno} we discuss the low-energy CP-odd interactions that are induced in the BSM models and the resulting EDMs of hadrons, nuclei, atoms, and molecules. In Sec.~\ref{results} we demonstrate how the pattern of EDMs of different systems can separate pure-$\bar\theta$ scenarios (which can arise either in  IR or UV solutions to the strong CP problem) from models where higher-dimensional flavor-diagonal CP-odd operators are relevant and an IR solution is needed. We conclude in Sec.~\ref{conclusions}.

\section{Corrections to $\bar\theta$}\label{theta-renorm}
In the SM, the strong CP phase that is invariant under anomalous chiral field redefinitions is:
\begin{align}
\bar\theta=\theta+{\rm arg\,det\,} y_u y_d\,,
\end{align}
where $y_{u,d}$ are the up- and down-type Yukawa matrices.
In BSM models, there can be additional terms in the invariant strong CP phase. There can also be new invariant phases that radiatively correct $\bar\theta$. We will see examples of both below. 

In perturbation theory, a radiative correction to ${\rm arg\,det\,} y$ shifts $\bar\theta$ at one loop by
\begin{align}
\label{eq:argdetimtr}
\Delta\bar\theta ={\rm arg\,det\,} (y+\delta y)-{\rm arg\,det\,}y &= {\rm arg\,det\,} (1 + y^{-1} \delta y) \approx {\rm Im\,tr\,}(y^{-1} \delta y),
\end{align}
where $y \equiv y_u y_d$.

\subsection{Corrections to $\bar \theta$ in SMEFT}

In SMEFT, there are quadratic divergences in the one-loop radiative contribution to $\bar\theta$ arising from higher-dimensional CP-violating operators. These terms can be used to estimate the correction to $\bar\theta$ in a systematic computation, as long as the Wilsonian momentum cutoff is taken sufficiently below the scale at which the SMEFT operators become strongly coupled. Typically, this is a conservative estimate. More generally, the quadratic divergences can be taken simply as an indicator that models which generate hadronic CP-violating SMEFT operators typically also have sizeable new threshold corrections to $\bar\theta$. The explicit calculation was performed in Ref.~\cite{deVries:2018mgf} and here we just list the result. 

We define the dimension-six operators through 
\begin{equation}
\mathcal L_{\rm SMEFT}= \mathcal L_{\rm SM} + \frac{1}{\Lambda^2} \sum_i c_i \mathcal O_i \,,
\end{equation}
in terms of the operators in Table~\ref{fig:optable}.  
The one-loop quadratic divergence in $\bar\theta$ from dimension-six CP-violating SMEFT operators is~\cite{deVries:2018mgf}
\begin{align}
16 \pi^2 \delta \bar \theta \sim & 16 \pi^2 \Big(  \frac{2}{g_s^2} c_{H\tilde G}  - \frac{9}{2 g_s} c_{\tilde G} \Big) \nonumber \\
& + \Im\, \Tr [Y_d^{-1} (3 c_{dH}  + g^\prime c_{dB} - 18 g c_{dW} - 16 g_s c_{dG} ) ] \nonumber \\
& + \Im\, \Tr [Y_u^{-1} (3 c_{uH}  -5 g^\prime c_{uB} - 18 g c_{uW} - 16 g_s c_{uG} ) ] \nonumber \\
& + \Im\, \Tr [(Y_d^{-1} Y_u + Y_d^\dagger (Y_u^\dagger)^{-1}) c_{Hud}] \nonumber \\ 
& + \Im [ 2 c_{lequ(1)}^{mnij} Y_e^{\dagger nm} (Y_u^{-1})^{ji} - 2 c_{ledq}^{*mnij} Y_e^{mn}(Y_d^{-1})^{ij}] \nonumber \\
& + \Im [(6 c_{quqd(1)}^{mnij} + c_{quqd(1)}^{inmj} + \frac{4}{3} c_{quqd(8)}^{inmj}) (Y_u^{\dagger nm}(Y_d^{-1})^{ji} + Y_d^{\dagger ji}(Y_u^{-1})^{nm} )]\,.
\label{eq:thetacorr}
\end{align}
Absent an infrared relaxation of $\bar\theta$, e.g.~by the Peccei-Quinn mechanism, naturalness requires $\abs{\delta \bar \theta} \lesssim 10^{-10}$, implying a stringent bound on the combination of Wilson coefficients in Eq.~\eqref{eq:thetacorr}.

\begin{table}[t!]
\begin{center}
\begin{tabular}{c | c || c | c}
$\op_{uH}$ & $H^\dagger H \overline{Q_{Li}} \tilde H u_{Rj}$ &
$\op_{dH}$ & $H^\dagger H \overline{Q_{Li}} H d_{Rj}$ \\
$\op_{dG}$ & $\overline{Q_{Li}} \sigma^{\mu\nu} T^a d_{Rj} H G^a_{\mu\nu}$ &
$\op_{dW}$ & $\overline{Q_{Li}} \sigma^{\mu\nu} d_{Rj} \tau^a H W^a_{\mu\nu}$ \\
$\op_{dB}$ & $\overline{Q_{Li}} \sigma^{\mu\nu} d_{Rj} H B_{\mu\nu}$ &
$\op_{uG}$ & $\overline{Q_{Li}} \sigma^{\mu\nu} T^a u_{Rj} \tilde H G^a_{\mu\nu}$ \\
$\op_{uW}$ & $\overline{Q_{Li}} \sigma^{\mu\nu} u_{Rj} \tau^a \tilde H W^a_{\mu\nu}$ &
$\op_{uB}$ & $\overline{Q_{Li}} \sigma^{\mu\nu} u_{Rj} \tilde H B_{\mu\nu}$ \\
$\op_{Hud}$ & $i \tilde H^\dagger D_\mu H \overline{u_{Ri}} \gamma^\mu d_{Rj}$ &
$\op_{quqd(1)}$ & $\epsilon^{ef} \overline{Q^e_{Li}} u_{Rj} \overline{Q^f_{Lk}} d_{Rl}$ \\
$\op_{quqd(8)}$ & $\epsilon^{ef} \overline{Q^e_{Li}} T^a u_{Rj} \overline{Q^f_{Lk}} T^a d_{Rl}$ &
$\op_{lequ(1)}$ & $\epsilon^{ef} \overline{L^e_{Li}} e_{Rj} \overline{Q^f_{Lk}} u_{Rl}$ \\
$\op_{ledq}$ & $\overline{L_{Li}} e_{Rj} \overline{d_{Rk}} Q_{Ll}$ &
$\op_{H \tilde G}$ & $H^\dagger H G^a_{\mu\nu} \widetilde G^{a \mu\nu}$ \\
$\op_{\tilde G}$ & $f^{abc} G^{a\mu}_{~~\nu} G^{b\nu}_{~~\rho} \widetilde G^{c\rho}_{~~\mu}$
\end{tabular}
\end{center}
\caption{Dimension-six SMEFT operators in the basis of Ref.~\cite{Grzadkowski:2010es} which contribute to the renormalization of $\bar \theta$ at one loop. \label{fig:optable}}
\end{table}

Now we  consider specific UV completions of the SM where the threshold corrections to $\bar\theta$ and the predictions for numerous EDM observables can be computed explicitly. We confirm the EFT expectation that whenever a dimension-six operator is introduced that appears in Eq.~\eqref{eq:thetacorr}, the dominant source of CP violation should be a $\bar \theta$ term much larger than the present bound, unless it is relaxed by a Peccei-Quinn mechanism.

Axion models~\cite{Peccei:1977ur,Peccei:1977hh,Wilczek:1977pj,Weinberg:1977ma, Kim:1979if,Shifman:1979if,Dine:1981rt,Zhitnitsky:1980tq} provide the standard IR solution to the strong CP problem. The QCD axion nonlinearly realizes an approximate chiral $U(1)_\text{PQ}$ symmetry, with decay constant $f_a \gtrsim 10^9\, \text{GeV}$. The axion $a$ modifies the CP-odd Lagrangian term for the gluons, $\mathcal L \rightarrow (\bar\theta + a/f_a) G \widetilde{G}$, so that the effective value of $\bar\theta(a)$ is modified by the axion vev. 
Nonperturbative QCD effects explicitly break $U(1)_\text{PQ}$ and generate a periodic potential for the axion that is minimized at $ \bar\theta + a/f_a = 0$, thus setting the effective CP-violating phase to zero in the vacuum of the axion model, $\langle \bar\theta(a) \rangle \equiv \bar\theta + \langle a \rangle / f_a = 0$. Additional sources of $U(1)_\text{PQ}$ violation can shift the minimum of the axion potential to a nonzero value of $\langle \bar\theta(a) \rangle$. To ensure a sufficiently small $|\bar\theta| < 10^{-10}$, the $U(1)_\text{PQ}$ global symmetry must be nearly exact, broken only by QCD effects or by very high-dimension irrelevant operators (often constrained to be higher than dimension-10). This is referred to as the axion quality problem. A handful of models are known to protect $U(1)_\text{PQ}$ to a sufficient degree without fine-tuning the coupling constants, often as an ``accident'' of some other structure in the ultraviolet theory~\cite{Chun:1992bn,Randall:1992ut,Cheng:2001ys,DiLuzio:2017tjx,Fukuda:2017ylt,Lillard:2017cwx,Lillard:2018fdt,Choi:2020vgb,Ardu:2020qmo,Nakai:2021nyf,Darme:2021cxx}.

\subsection{Case I:  Scalar Leptoquark Models}\label{sec:LQ}
We start our analysis in a very simple SM extension, a model of scalar leptoquarks (LQs). Recently these models have  been investigated extensively  due to anomalies in $b\rightarrow sll$ transitions and the muon anomalous magnetic moment. We consider here the $R_2$ scalar LQ \cite{Dorsner:2016wpm} that carries SM gauge quantum numbers $(3,2,7/6)$. Apart from interactions among LQs and LQ-Higgs interactions that conserve CP, there are potentially CP-violating interactions between LQs and the SM fermions. We focus on couplings between quarks and charged leptons given by
\bea\label{LQ_int}
\vL_Y^{(R_2)}=R_2^I\left( \bar u_R x_{RL} \epsilon_{IJ} L^J+ \bar Q^I x_{LR}^\dagger e_R\right) +{\rm h.c.}\,,\label{eq:R2lag}
\eea
where $x_{RL,LR}$ are  $3\times 3$ flavor matrices and $I,J$ denote $SU(2)$ indices. The presence of two interactions coupling to both left- and right-handed quarks and leptons leads to a rich EDM phenomenology \cite{Dekens:2018bci,Fuyuto:2018scm}. The LQ doublet describes two states with charges $(5/3)e $ and  $(2/3)e$ and we consider a common mass for these states $m_{R_2}$.  
 
The relative phase in the LQ couplings lead to a threshold correction to $\bar \theta$. That is, even if $\bar\theta=0$, it is renormalized at one loop by phases in the $x$ couplings. Closing the lepton lines gives the following UV-divergent contribution 
\begin{align}
\Delta\bar\theta = \frac{c}{8\pi^2} {\rm Im\,Tr\,} [y_u^{-1}(x_{RL}y_e^\dagger x_{LR}) ]\log\left(\frac{\Lambda}{m_{R_2}}\right) \,,
\label{eq:deltathetaLR}
\end{align}
where $\Lambda$ is a momentum cutoff and $c$ is an order one constant. We see the appearance of a second reparametrization-invariant phase, 
\begin{align}
-{\rm arg\,det\,}y_u-{\rm arg\,det\,}y_e+{\rm arg\,det\,}x_{RL}+{\rm arg\,det\,}x_{LR}\,.
\end{align}
While this contribution is suppressed by the lepton Yukawa, it can be enhanced by $1/y_u$ for first-generation quarks. The same couplings also give rise to a correction to the electron mass 
\begin{align}
\Delta m_e &\sim  \frac{1}{8\pi^2} (x_{RL}^{eq}m_{q}x_{LR}^{q e})\log\left(\frac{\Lambda}{m_{R_2}}\right),
\end{align}
and on naturalness grounds we require $\Delta m_e\lesssim m_e$. This says, for example, that 
\begin{align}
x_{RL}^{et}x_{LR}^{t e}\log(\Lambda/m_{R_2}) \lesssim 10^{-4}.
\end{align}
 However, it places essentially no constraint on first-generation couplings.

The phenomenology of EDMs in leptoquark models was discussed in detail in Refs.~\cite{Dekens:2018bci, Fuyuto:2018scm}. The LQ-fermion interactions lead to dimension-six operators below the scale of the LQ masses. Tree-level diagrams induce \cite{He:1992dc,Barr:1992cm}
\begin{eqnarray}
\mathcal L &=& \frac{c^{abcd}_{lequ (1)}}{\Lambda^2}(\bar L_a^I  e_{R_b}) \epsilon_{IJ}(\bar Q_c^J  u_{R_d})+ \frac{c^{abcd}_{lequ (3)}}{\Lambda^2}(\bar L_a^I  \simu e_{R_b})\epsilon_{IJ}\,(\bar Q_c^J  \sigma_{\mu\nu} u_{R_d})\,,
\label{eq:EFT4fermion}
\eea 
where 
\begin{equation}
\frac{c^{abcd }_{lequ (1)}}{\Lambda^2}=4\,\frac{c^{abcd }_{lequ(3)}}{\Lambda^2}=\frac{\left(x_{LR}\right)^{bc}\left(x_{RL}\right)^{da}}{2 m_{R_2}\sq}\,.
\end{equation}

Note that closing the lepton lines into a loop, the first operator contributes to the renormalization of $\bar\theta$ in SMEFT, cf. Eq.~(\ref{eq:thetacorr}), which correctly matches the qualitative structure of the correction in the full LQ model, Eq.~(\ref{eq:deltathetaLR}). In this model we have a direct correspondence between the EFT and full theory computations of the correction to $\bar\theta$. In the other models we consider, the SMEFT estimates will be conservative, with more important corrections to $\bar\theta$ arising already at tree level in the UV theory. The second operator also contributes to $\bar \theta$ but only at two-loop order. This is reflected by the mixing of $c^{abcd }_{lequ (1)}$ and $c^{abcd }_{lequ (3)}$ at one-loop order in QCD.

At the threshold of the top and charm quark, the four-fermion operators induce dimension-seven CP-violating lepton-gluon operators. We only consider the electron interactions
\bea\label{eG}
\vL_{eG} = C_{eG}\,\al_s\bar e \, i\g_5 e \, G^a_{\mu\nu}G^{a\, \mu\nu}+ C_{e\tilde G}\,\frac{\al_s}{2}\bar e  e \, G^a_{\mu\nu}G^{a}_{\al\bt} \epsilon^{\al\bt\mu\nu}\,,
\eea
where 
\bea
C_{eG} = -\frac{2}{3}C_{e\tilde G}=  \sum_{q=c,t}\frac{1}{24\pi}\frac{{\rm Im}\, c_{le qu(1)}^{eeqq}}{\Lambda^2\,m_q}
\,.\eea

Electric dipole moments of quarks and leptons and chromo-electric dipole moment of quarks are induced at the one-loop level. Focusing on first-generation quarks and leptons we define the operators\footnote{The dipole operators can be related to the coefficients in Eq.~\ref{eq:thetacorr}. For instance, $d_{u} = (e \sqrt{2} v/\Lambda^2)\,\mathrm{Im}\,\left( c^{11}_{uB}/g' + c^{11}_{uW}/g\right)$ and $\tilde d_{u} = -(\sqrt{2} v/\Lambda^2)\,\mathrm{Im} \left( c^{11}_{uG}/g_s\right)$\,. The electron EDM does not appear in Eq.~\eqref{eq:thetacorr} as it only renormalizes $\bar \theta$ at the three-loop level.}
\begin{eqnarray}
\vL_{\rm dipole}&=&-\frac{d_e}{2} \,\bar e\,\sigma^{\mu\nu}i\gamma^5\,e\,F_{\mu \nu} -\frac{d_u}{2} \,\bar u\,\sigma^{\mu\nu}i\gamma^5\,u\,F_{\mu \nu}  - \frac{g_s  \tilde d_u}{2} \,\bar u\,\sigma^{\mu\nu}i\gamma^5 t^a\,u\,G_{\mu \nu}^a\,,
\label{eq:EFTdipoles}
\end{eqnarray}
where the one-loop expressions are given by
\begin{eqnarray}\label{oneloopLQ}
d_e &=& \frac{e}{(4\pi)^2} \sum_{q=u,c,t} \frac{N_c m_q}{6}\, \mathrm{Im}\left[\frac{\left(x_{LR}\right)^{eq}\left(x_{RL}\right)^{qe}}{ m_{R_2}\sq}\right] + \mathcal O\left(\frac{m_q^2}{m_{R_2}\sq}\right)\,,\nonumber\\
d_u &=& -\frac{e}{(4\pi)^2} \sum_{l=e,\mu,\tau} \frac{2 m_l}{3}\, \mathrm{Im}\left[\frac{\left(x_{LR}\right)^{lu}\left(x_{RL}\right)^{ul}}{ m_{R_2}\sq}\right] + \mathcal O\left(\frac{m_l^2}{m_{R_2}\sq}\right)\,,\nonumber\\
\tilde d_u &=& -\frac{1}{(4\pi)^2} \sum_{l=e,\mu,\tau} \frac{m_l}{2}\, \mathrm{Im}\left[\frac{\left(x_{LR}\right)^{lu}\left(x_{RL}\right)^{ul}}{ m_{R_2}\sq}\right] + \mathcal O\left(\frac{m_l^2}{m_{R_2}\sq}\right)\,.
\end{eqnarray}
In particular the electron EDM can get large contributions from internal top quarks.

If we focus on LQ couplings for first-generation quarks and leptons only, the largest contributions are to CP-odd four-fermion operators while the dipole operators are loop suppressed. For electron--top LQ interactions this is no longer the case, as the electron-gluon operators also arise at one loop and the electron EDM is enhanced by $m_t/m_e$. The resulting EDM phenomenology will be discussed in Sect.~\ref{EDMpheno}.

\subsection{Case II: the Minimal Supersymmetric Standard Model}
In the MSSM, the invariant strong CP phase is
\begin{align}
\bar\theta=\theta+{\rm arg\,det\,} y_u y_d+3\arg m_{\tilde g} + 3\arg v_u v_d
\end{align}
It receives new tree level contributions from the last two terms. It also receives radiative corrections, for example, from one-loop contributions to the quark masses or the gluino mass that are sensitive to phases in the $A$-terms and the $\mu$ term~\cite{Dine:1993qm}. 
Rather than studying the full set of one-loop corrections to $\bar\theta$, we restrict our attention to those contributions that are proportional to $\alpha_s$, as this is sufficient to generate a nontrivial pattern of electric dipole moments.  

At one-loop order in $\alpha_s$, the Yukawa couplings $y_{u,d}$ receive two types of corrections.
The first type is proportional to $A_{u,d}$, and in a highly-simplified limit the correction to $\bar\theta$ is
\begin{align}
\Delta\bar\theta & \approx \frac{\alpha_s}{3\pi} \frac{1}{m_\text{soft}^2} {\rm Im\,Tr\,} [ y_u^{-1}(m_{\tilde g}^\dagger A_u)  ] +(u\leftrightarrow d).
\end{align}
Here we  work in a flavor basis where the gluino-quark-squark couplings are real and proportional to the identity in flavor space, and for simplicity we take the limits where the squarks are degenerate at $m_\text{soft}$, the gluino has similar mass $|m_{\tilde g}|^2 \approx m_\text{soft}^2$, and the $A$-terms are treated in the insertion approximation. 
We see that if the gluino mass is real, and in the limit of exact proportionality $A_f\propto y_f$,  the corrections to $\bar\theta$ vanish. However, if we allow  for deviations from exact proportionality, $\Delta\bar\theta$ is in general nonzero because of phases in the Yukawas, even if the gluino phase vanishes at tree level. Furthermore, if we allow deviations from exact degeneracy of the squark masses,  the trace above will include insertions of flavor-changing squark soft masses $\delta \tilde m_{f f'}^2$, e.g.:
\begin{align}
\left(\Delta\bar\theta \right)_A &\simeq \frac{\alpha_s}{18\pi} \frac{1}{m_\text{soft}^6} {\rm Im\,Tr\,} [ y_u^{-1}m_{\tilde g}^\dagger( \delta \tilde m^2)  A_u (\delta \tilde m^2)  ] +(u\leftrightarrow d).
\end{align}
This correction can shift $\bar\theta$ even in the limit of exact $A$~term--Yukawa proportionality.

A second contribution to the quark masses comes from the $\mu$ term in the superpotential, which induces cubic scalar interactions $\mathcal L \supset \mu^\star [\tilde u_R y_u \tilde u_L H_d^{0 \star}  + (d \leftrightarrow u) ] + h.c$. With the $H_{u,d}^0 \rightarrow v_{u,d} / \sqrt{2}$ vev insertion, 
\begin{align}
\delta m_u^{ij} \approx - \frac{\alpha_s}{3\pi} \mu^\star m_g^\star \frac{v_d^\star }{\sqrt{2}} y_u^{ij} \frac{1}{m_\text{soft}^2},
\end{align}
where again we simplify the loop integrals by setting the squark and gluino propagator masses to a universal $m_\text{soft}$. An analogous expression for $\delta m_d^{ij}$ is found by applying $(u \leftrightarrow d)$ above. Each correction $\delta y_q$ is strictly proportional to $y_{q}$, so the new contribution to $\bar\theta$ is
\begin{align}
\left(\Delta\bar\theta \right)_\mu &\simeq \text{Im\,Tr}(y_u^{-1} \delta y_u + y_d^{-1} \delta y_d ) = - \frac{\alpha_s}{\pi} \,\text{Im}\!\left[ \left( \frac{v_d^\star}{v_u} + \frac{v_u^\star}{v_d} \right) \frac{\mu^\star m_g^\star}{m_s^2} \right] .
\end{align}
Note that in the limit of large $\tan \beta \equiv |v_u/v_d|$, the assumption $\delta y_u \ll y_u$ underpinning the Eq.~\eqref{eq:argdetimtr} approximation can become invalid if $\mu \tan\beta \alpha_s  / \pi \gtrsim \mathcal O(m_s)$, so a more careful analysis is required in that regime~\cite{Ellis:2008zy}.

From the gluino mass correction, $\Delta \bar\theta \approx 3 \, \text{Im}(m_g^{-1} \delta m_g)$, with similar assumptions for the particle masses and couplings, we obtain:
\begin{align}
\left(\Delta\bar\theta\right)_{v^2} &= -\frac{3\alpha_s}{16\pi}  \frac{v_u^2}{m_\text{soft}^2} \text{Im} \left(m_{\tilde g}^{-1} \, \text{Tr} (A_u y_u^\dagger)  \right) +(u\leftrightarrow d).
\end{align}

In each case, we see the appearance of  additional  phases that can correct $\bar\theta$. The relevant reparametrization-invariant phases can be read off directly from each of the corrections (and in cases involving vevs note that the soft SUSY-breaking $B$ term transforms like $v_u^*v_d^*$). 
If there is a large hierarchy between the weak scale and the susy-breaking scale, the correction to the gluino mass phase  can be suppressed below $10^{-10}$; however, this suppression does not affect the correction to the quark mass phases. 
To remove all the large sources of $\Delta \bar\theta$ in the MSSM, additional strong  assumptions must be made. For example, in soft supersymmetry breaking universality, the Higgs vevs and gaugino masses are taken to be real, and the $A$ terms are taken to be exactly proportional to the Yukawa matrices.

In the MSSM with R-parity conservation, there are no tree level contributions to SMEFT operators, but the dipole operators $c_{uB,uW,uG,dB,dW,dG}$ are induced at one loop. They appear in Eq.~\eqref{eq:thetacorr}; closing the gauge-boson loop corresponds to a two-loop contribution to $\bar \theta$. In the full model there are larger, one-loop corrections to $\bar\theta$ from diagrams with the gauge boson removed entirely, not to mention tree-level contributions. So in this case the SMEFT estimate  is likely conservative. Nevertheless, it is convenient as a diagnostic tool: the appearance of any dimension-six operators appearing in Eq.~\eqref{eq:thetacorr}, which can be verified by low-energy experiments as discussed in detail below, suggests large corrections to $\bar \theta$. 

The quark EDMs and CEDMs ($d_q$ and $\tilde d_q$, respectively) are defined as in Eq.~\eqref{eq:EFTdipoles} 
\begin{equation}
\mathcal L =  - \frac{d_q}{2} \bar q \sigma^{\mu\nu} i \gamma^5 q\,F_{\mu\nu} - \frac{ g_s \tilde d_q}{2} \bar q \sigma^{\mu\nu} i \gamma^5  t^a q\,G_{\mu\nu}\,,
\end{equation}
for $q=\{u,d\}$.
Complex phases in the gluino mass  and in the fermion--gluino interaction generate an EDM and CEDM at one-loop order, proportional to $\alpha_s$. For the EDM, the relevant diagram features a virtual gluino and squark, $q \rightarrow (\tilde g + \tilde q') \rightarrow q$, with a photon coupled to the virtual squark. 
Here the deviations from exact degeneracy in the squark masses become important, so we will treat them with a degree of generality, including off-diagonal $(\text{mass})^2$ couplings in the flavor basis. In the absence of any generation-changing masses, the first-generation up and down squarks (in the flavor basis) have mass matrices given approximately by~\cite{Ibrahim:1997gj}
\begin{align}
M_{\tilde u}^2 &\simeq  \left(
\begin{array}{c c } M_{\tilde Q}^2 + \mathcal O(m_Z^2 \cos 2\beta)  		& m_u(A_u^\star  - \mu \cot \beta) \\ m_u (A_u  - \mu^\star \cot \beta) 	& M_{\tilde U}^2 +   \mathcal O(m_Z^2 \cos 2\beta)   \end{array} \right) \approx m_\text{soft}^2 \left( \begin{array}{c c } 1 & \epsilon_u^\star \\ \epsilon_u & 1 + \Delta_u \end{array} \right) ,
\\
M_{\tilde d}^2 &\simeq  \left(
\begin{array}{c c } M_{\tilde Q}^2 + \mathcal O(m_Z^2 \cos 2\beta)  		& m_d(A_d^\star  - \mu \tan \beta) \\ m_d (A_d  - \mu^\star \tan \beta) 	& M_{\tilde D}^2 +   \mathcal O(m_Z^2 \cos 2\beta)   \end{array} \right)  \approx m_\text{soft}^2 \left( \begin{array}{c c } 1 & \epsilon_d^\star \\ \epsilon_d & 1 + \Delta_d \end{array} \right) ,
\end{align}
where the $A_{u,d}$ above refer to the $(1,1)$ components of the respective $3\times 3$ matrices. 
Above, we parametrize the mass matrices with a real $\Delta_q$ and complex $\epsilon_q$ for each $q=u,d$. 
Ignoring the $\mathcal O(m_Z^2)$ diagonal entries, we take the $(1,1)$ (i.e. left-left) components of the two matrices to be approximately equal to $m_\text{soft}^2 \approx M_{\tilde Q}^2$.

 In the approximately-degenerate limit assumed for the squark masses in our calculation of the one-loop corrections to $\Delta \bar\theta$, $|\Delta_q | \ll \mathcal O(1)$. Even with exactly degenerate soft SUSY-breaking masses $M_{\tilde Q}^2 = M_{\tilde U}^2 = M_{\tilde D}^2$, however, the $\mathcal O(m_Z^2)$ diagonal terms from electroweak symmetry breaking introduce some degree of mass splitting. Absent any fine cancellations between the electroweak corrections and the soft masses, we anticipate a lower limit on the magnitude of $\Delta$, $|\Delta_q | \gtrsim \mathcal O( \cos 2\beta \,m_Z^2 / m_\text{soft}^2)$.
The off-diagonal $\epsilon_q$, on the other hand, are suppressed by factors of $m_q / m_\text{soft}$, so we may assume $|\epsilon | \ll |\Delta |$ for the first-generation quarks.

Following~\cite{Ellis:2008zy,Ibrahim:1997gj}, the contribution from the gluino loop diagram to the quark EDM $d_q$ is given by
\begin{align}
\frac{d_q}{e} = - \frac{2 \alpha_s}{3 \pi} \sum_{j = 1,2} \text{Im}\!\left( U_{q 2 j} U^\star_{q 1 j} e^{- i \phi_3} \right) \frac{|m_{\tilde g}|} {M_{\tilde q_j}^2 } Q_{\tilde q} \, B\!\left( \frac{|m_{\tilde g}|^2 }{M_{\tilde q_j}^2 } \right) ,
\end{align}
where $\phi_3 = \text{arg}\, m_{\tilde g}$ is the phase of the gluino mass; the   $U_q$ diagonalize the mass matrices via $U_q^\dagger M_{\tilde q}^2 U_q = \text{diag}\left( M_{\tilde q_1}^2, M_{\tilde q_2}^2  \right)$, with mass eigenvalues $M_{\tilde q_j}^2$; 
$Q_{\tilde q}$ is the  squark electric charge;
and where
\begin{align}
B(r) = \frac{1 + r + \frac{2 r  }{1 - r} \ln r }{2 (1- r)^2} ,
&&
B(1) = \frac{1}{6}.
\end{align}
Expanding to linear order in $\epsilon/\Delta$, we find $(U_q)_{11} = (U_q)_{22} \simeq 1$, and $(U_q)_{12} = - (U_q)_{21}^\star = \epsilon^\star / \Delta$. Similarly, the mass eigenvalues are $M_{\tilde q_1}^2 \simeq m_\text{soft}^2 (1 + \mathcal O(\epsilon^2))$ and $M_{\tilde q_2}^2 \simeq m_\text{soft}^2 (1 + \Delta + \mathcal O(\epsilon^2))$.
Specializing to the $|m_{\tilde g}^2 | \approx m_\text{soft}^2$ case, we find
\begin{align}
\frac{d_q}{e} &\approx - Q_q \frac{\alpha_s}{4\pi} \frac{|\epsilon_q|}{m_\text{soft}} \sin\left( \phi_3 - \text{arg} \,\epsilon_q \right) \times F(\Delta), \\
F(\Delta) &\equiv \frac{4}{9 \Delta^4}\left[ \Delta^3- 3\Delta^2 -6 \Delta  + 6 (1+\Delta) \log (1+\Delta) \right]\,.
\end{align}
In the limit of degenerate squark masses, $\Delta \rightarrow 0$, the function $F$ approaches a constant, $F(0) = 2/9$.
Recalling that the typical size of $\epsilon$ is $|\epsilon| \sim m_q / m_\text{soft}$, the overall contribution to the $q = u,d$ dipole moment scales as $m_q / m_\text{soft}^2$.

At one-loop order, the quark chromo-EDM is proportional to the same $\text{Im}(e^{i \phi_3} \epsilon^\star)$ combination of phases, with the same dependence on $U_q$; the only major difference is the existence of a second type of diagram,
where the external gluon couples to the virtual gluino rather than the squark, so that the gluino loop diagram induces a CEDM of
\begin{eqnarray}
\tilde d_q\approx  \frac{\alpha_s}{4\pi} \frac{|\epsilon_q|}{m_\text{soft}}  \sin\left( \phi_3 - \text{arg} \,\epsilon_q \right) \times G(\Delta)\,,
\end{eqnarray}
where
\begin{eqnarray}
G(\Delta) = \frac{1}{18 \Delta^4}\left[ 19\Delta^3 + 78 \Delta^2 + 48 \Delta  - 6 (1+\Delta)(8+9\Delta) \log (1+\Delta) \right]\,.
\end{eqnarray}
Here we have again specialized to $m_{\tilde g}^2 \approx m_\text{soft}^2$ in order to express the result simply in terms of $\Delta$.
In the limit $\Delta \rightarrow 0$, $G(0) = 5/18$ remains finite.

By turning on additional phases in the electroweak chargino sector, the EDMs and CEDMs receive further contributions, but as these are proportional to $g^2$ or $g'^2$, they can be considered small compared to the $\alpha_s$ corrections listed above unless the gluino phase is tuned so that $\text{Im}(e^{i \phi_3} \epsilon_q^\star)$ is small.

\subsection{Case III:  the P-symmetric minimal left--right symmetric model}
Left--right symmetric models have an extended gauge symmetry $SU(3)_c \times SU(2)_L \times SU(2)_R\times U(1)_{B-L}$ \cite{Pati:1974yy, Mohapatra:1974hk, Senjanovic:1975rk} . The left- and right-handed fermions are fundamental representations of the $SU(2)$ groups, and right-handed neutrinos appear automatically. The minimal left--right symmetric model (mLRSM) includes a minimal scalar sector containing one scalar bidoublet and two triplets \cite{Dekens:2014ina}. Vacuum expectation values of these scalar fields then break the extended gauge symmetry to the SM group at some high-energy scale. This leads to massive right-handed electroweak gauge and scalar bosons with masses above a few TeV to avoid phenomenological constraints. Full details of the model can be found in many places in the literature, see e.g.~\cite{Pati:1974yy, Mohapatra:1974hk, Senjanovic:1975rk,Maiezza:2010ic,Dekens:2014jka}. 

Left--right models can have an exact symmetry between left- and right-handed fermions at high scales. One way of doing so is by considering a generalized parity ($P$) symmetry. In the case of exact $P$ symmetry the QCD $ \theta$ term is explicitly forbidden in the microscopic theory. However, there is an explicit CP-violating phase $\delta_2$ in the Higgs potential, 
\begin{align}
\alpha_2\{e^{i\delta_2}[{\rm Tr}(\phi \tilde\phi^\dagger){\rm Tr}(\Delta_R \Delta_R^\dagger)+{\rm Tr}(\phi^\dagger \tilde\phi){\rm Tr}(\Delta_L \Delta_L^\dagger)]+h.c.\}.
\end{align} 
Here $\phi$ is the Higgs bidoublet and $\Delta_{L,R}$ are Higgs triplets. $\delta_2$ is responsible for spontaneous CP violation in the Higgs bidoublet, $\langle\Phi\rangle={\rm diag}(v_1,e^{i\alpha}v_2)$, with $\sin(\alpha)\propto\sin(\delta_2)$~\cite{Kiers_2005,Maiezza:2014ala}.  Thus the quark mass matrices, and $\bar\theta$, obtain a tree-level phase of order $\delta_2$, unsuppressed by any ratio of scales.

In left--right symmetric model, CP-odd flavor-diagonal operators are induced already at tree-level. They arise from the exchange of heavy $W_R$ and  scalar fields. The contributions from the latter lead to four-fermion operators that scale with SM Yukawa couplings and are thus suppressed for light fermions most relevant for EDMs. In addition, to avoid producing too large flavor-changing neutral currents the scalar fields must be relatively heavy whereas the $W_R$ boson could be lighter. We therefore focus on contributions from the tree-level exchange of $W_R$ bosons. The tree-level matching leads to a single SMEFT operator right below the mass of the $W_R$ boson (see Ref.~\cite{Dekens:2014jka} for a derivation)
\bea
\vL_{6,\rm{mLRSM} } &=&   \frac{c^{ij}_{Hud}}{\Lambda^2} i \tilde{\vp}^{\dagger} D_{\mu} \vp \, \bar{u}^i_R \gamma^\mu  d^j_R 
+  \mathrm{h.c.}\, ,
\label{dim6edms}
\eea
where 
\begin{eqnarray}\label{eq:mHmatching}
\frac{c^{ij}_{Hud}}{\Lambda^2}  &=&\frac{g_R^2 }{ m_{W_R}^2} \frac{\xi e^{i\alpha}}{1+\xi^2}V_{R,\,ij}\,,
\end{eqnarray}
in terms of a gauge coupling $g_R=g$, the mass of the right-handed gauge boson $m_{W_R}$ and the parameter $\xi$ related to the ratio of vevs appearing in the model. In this section, we set $V_R = V_{\rm CKM}$ which is a leading-order expression in the limit  $\xi\sin\al\to 0$  \cite{Senjanovic:2014pva}. Corrections can be systematically included but do not change the qualitative conclusions. 

The dimension-six operator $c^{ij}_{Hud}$ appears in Eq.~\eqref{eq:thetacorr} and thus indicates a renormalization of $\bar \theta$ at the matching scale. In this case, the EFT argument is conservative as the renormalization already appears at tree level. Nonetheless from a low-energy perspective the appearance of $c^{ij}_{Hud}$ signifies the need to account for the large $\bar \theta$ term.

Below the electroweak scale, after integrating out the electroweak gauge bosons, we obtain four-quark operators 
\begin{equation}
\label{eq:Leff}
\tilde{\cal L}_{\rm eff}  =  
 - \sum^{2}_{a=1} \,\left(  C^{ij\, lm}_{a\, LR} \mathcal O^{ij\, lm}_{a\, LR} + C^{{ij\, lm}\,*}_{a\, LR} \big(\mathcal O^{ij\, lm}_{a\, LR}    \big)^\dagger  \right)\ ,      
\end{equation}
where
\begin{eqnarray}\label{eq:4quark1}
\mathcal O^{ij\, lm}_{1\, LR} = \bar d^m \gamma^\mu P_L u^l \, \bar u^i \gamma_\mu P_R d^j\ , \qquad  \mathcal O^{ij\, lm}_{2\, LR}  =\bar d_\al^m   \gamma^\mu P_L u_\bt^l \, \bar u_\bt^i  \gamma_\mu P_R d_\al^j\ ,
\end{eqnarray}
and  $\alpha$ and $\beta$ are color indices. At the electroweak scale we have
\begin{eqnarray}\label{eq:4q2}
C^{ij\, lm}_{1\, LR}(m_W)  =\frac{V^*_{lm}c^{ij}_{Hud}}{\Lambda^2}\ , \qquad C^{ij\, lm}_{2\, LR}(m_W) = 0 \ .
\end{eqnarray}
The CP-odd four-quark operators depend on the same phase as the induced correction to $\bar \theta$. 

We focus on operators involving just up, down, and strange quarks and find four relevant operators
\begin{eqnarray}\label{eq:4quark}
\mathcal L_{\textrm{EDM}} &=& - i \Big(   \textrm{Im}\, C^{ud\, ud}_{1\, LR} \, \bar d \gamma^\mu P_L  u \, \bar u \gamma_\mu P_R d   
+ \mathrm{Im}\, C^{ud\, ud}_{2\, LR} \,  \bar d_\alpha \gamma^\mu P_L  u_\bt \, \bar u_\bt \gamma_\mu P_R d_\alpha  \nonumber
 \\ & & +  \textrm{Im} \, C^{us\, us}_{1\, LR}  \,  \bar s \gamma^\mu P_L  u \, \bar u \gamma_\mu P_R s 
 +  \textrm{Im} \, C^{us\, us}_{2\, LR}  \, \bar s_\al \gamma^\mu P_L  u_\bt \, \bar u_\bt \gamma_\mu P_R s_\al - \text{h.c.} \Big) 
\end{eqnarray}
where the second operator is induced through QCD renormalization-group evolution \cite{Hisano:2012cc,Dekens:2013zca} 
\begin{eqnarray}
C_{1\, LR}^{ijlm}(3\, {\rm GeV})= 0.9\, C_{1\, LR}^{ijlm}(m_W)\ ,\qquad
C_{2\, LR}^{ijlm}(3\, {\rm GeV})=  0.4\,C_{1\, LR}^{ijlm}(m_W)+1.9\,C_{2\, LR}^{ijlm}(m_W)\ .
\end{eqnarray}

\section{EDM phenomenology}
\label{EDMpheno}
\subsection{CP-violating hadronic couplings}
As discussed above, models with CP violating couplings to the strong sector typically induce a large $\bar\theta$ term. For consistency with the neutron EDM the $\bar\theta$ term must be relaxed further in the infrared, presumably by a Peccei-Quinn mechanism. The pattern of higher-dimension CP-violating operators generated alongside $\bar\theta$ in each of the models survives, producing distinctive patterns of  electric dipole moments at lower energies. We now turn to the estimation of these EDM observables, a somewhat complicated task that involves hadronic, nuclear, atomic, and molecular physics. In the spirit of effective field theories, it is useful to perform the calculation in steps. We first discuss what hadronic or semi-leptonic CP-violating operators are generated from the various CP-violating operators at the quark-gluon level. In principle, many hadronic interactions are generated, but they can be organized in a systematic way in chiral perturbation theory. In this section, we give the main results and refer to Refs.~\cite{deVries:2012ab,deVries:2020iea} for more details.

{\bf Semi-leptonic operators.}~We begin with the semi-leptonic operators that describe CP-violating electron--quark and electron--gluon interactions. These operators are induced in the leptoquark model, cf.  Eq.~\eqref{eq:EFT4fermion} and \eqref{eG}. They further induce CP-odd electron-nucleon interactions that, in turn, induce atomic and molecular EDMs. The CP-violating electron-nucleon interactions take the form \cite{Dekens:2018bci} 
\bea\label{hadronicCPV} \vL & =&-\frac{G_F}{\sqrt{2}}\bigg\{\bar e i\g_5 e\, \bar N\left(C_S^{(0)}+\tau_3 C_S^{(1)}\right) N + \bar e e\, \frac{\partial_\mu}{m_N} \left[\bar N\left(C_P^{(0)}+\tau_3 C_P^{(1)}\right)S^\mu N\right]\nn\\
&&-4\, \bar e \sigma_{\mu\nu} e\, \bar N\left(C_T^{(0)}+\tau_3 C_T^{(1)}\right) v^\mu S^\nu N\bigg\}+\dots \,,
\eea
where $N
= (p\,\,n)^T$  is the non-relativistic nucleon doublet with mass $m_N$, velocity $v^\mu$, and the spin $S^\mu$ ($v^\mu =(1,\boldsymbol 0)$ and $S^\mu = (0,\, \boldsymbol
\sigma/2 )$ in the nucleon rest frame). The matching coefficients are given by
\bea \label{eq:matchCSP}
C_S^{(0)} &=& v\sq\left[\frac{\sigma_{\pi N}}{m_u+m_d}{\rm Im}\, C_{lequ}^{(1)\, eeuu}+\frac{16\pi}{9}(m_N-\sigma_{\pi N}-\sigma_s)C_{eG}\right]\,,\nn\\
C_S^{(1)}& =& v\sq\frac{1}{2}\frac{\dt m_N}{m_d-m_u}{\rm Im}\, C_{lequ}^{(1)\, eeuu}\,,\nn\\
C_P^{(0)} &=&-8\pi v^2(\Delta_u+\Delta_d) m_NC_{e\tilde G}\,,\qquad C_P^{(1)} = v\sq \frac{g_Am_N}{m_u+m_d}{\rm Im}\, C_{lequ}^{(1)\,eeuu}-8\pi v^2g_A m_N\frac{m_d-m_u}{m_u+m_d}C_{e\tilde G} \,,\nn\\
C_T^{(0)} &=& v^2(g_T^d+g_T^u){\rm Im}\, C_{lequ}^{(3)\, eeuu}\,,\qquad\,\,\, C_T^{(1)} \,\,= v^2(g_T^d-g_T^u){\rm Im}\, C_{lequ}^{(3)\, eeuu}\,.
\eea
in terms of the hadronic matrix elements  \cite{Airapetian:2006vy, Hoferichter:2015dsa, Abdel-Rehim:2016won, Borsanyi:2014jba} 
\begin{align}
\sigma_{\pi N} &= (59.1 \pm 3.5)\,\mathrm{MeV}\, ,& \sigma_s &= (41.1_{-10.0}^{+11.3})\,\mathrm{MeV} \, , &
\delta m_N &=( 2.32\pm0.17)\, \mathrm{MeV}\, ,\nn\\
  g_A &= 1.27\pm0.002\,,& \Delta_u &= 0.842\pm 0.012\,,& \Delta_d &= -0.427\pm 0.013\,.
\end{align}

{\bf Hadronic operators.}~More complicated are the purely hadronic operators such as the quark (chromo-)EDMs and four-quark operators. We begin with the analysis of quark EDMs, which are induced in the LQ scenario as well as the MSSM. Due to the explicit appearance of the electromagnetic field strength, quark EDMs mainly induce hadronic operators that contain explicit photons as well (operators without photons are suppressed by $\alpha_{\mathrm em}/\pi$). The most important operators are the nucleon EDMs, related to the quark EDMs by
\begin{eqnarray}
d_n(d_q) &=& g_T^u d_u + g_T^d d_d\,,\nonumber\\
d_p(d_q) &=& g_T^u d_d + g_T^d d_u\,.
\end{eqnarray}
where $g_T^u = -0.213 \pm 0.011$ and $g_T^d = 0.820 \pm 0.029$. These so-called tensor charges are obtained from lattice-QCD calculations \cite{Gupta:2018lvp} and have very small theoretical uncertainties. 

The quark chromo-EDMs also contribute to nucleon EDMs but there are no lattice-QCD calculations available at present. The neutron EDM was evaluated using QCD sum rules \cite{Pospelov_qCEDM,Hisano:2012sc} giving
\begin{equation}
d_n(\tilde d_q) = \tilde g_n \left(4 Q_d \tilde d_d - Q_u \tilde d_u\right)\,,
\end{equation}
where $\tilde g_n = (1\pm 0.5) 0.55 e/Q_u$. We express the proton EDM through a quark model relation 
\begin{equation}
d_p(\tilde d_q)  =  \tilde c_p \tilde g_n \left(- 4 Q_d \tilde d_u + Q_u \tilde d_d\right)\,,
\end{equation}
so that $d_p$ and $d_n$ depend on the same QCD matrix element $\tilde g_n$. We use $\tilde c_p = 1\pm 0.2$ to account for possible isospin breaking. These relations are valid only under a Peccei-Quinn mechanism, that is the expressions take into account the contribution from the induced $\bar \theta$ term. 

In addition to nucleon EDMs, the quark chromo-EDMs also induce CP-violating pion-nucleon interactions. The most important operators are given by 
\begin{equation}
\mathcal L = \bar g_0\,\bar N \tau \cdot \pi\,N + \bar g_1\,\bar N \pi_3\,N \,,
\end{equation}
in terms of the pion triplet $\vec \pi$. These couplings were evaluated with QCD sum rules as well, but come with rather large uncertainties \cite{Pospelov_piN}. Chiral perturbation theory can be used to obtain some further insight in these matrix elements. Through a chiral transformation it can be shown that the matrix elements connecting $\bar g_{0,1}$ to quark chromo-EDMs are related to matrix elements connecting meson and baryon mass corrections to CP-even quark chromo-magnetic dipole moments \cite{deVries:2016jox}. This by itself does not help as the matrix elements of  quark chromo-magnetic dipole moments are also poorly known. However, Ref.~\cite{Seng:2018wwp} argued that the unknown matrix elements (called the `direct' contributions) are small compared to so-called vacuum-alignment pieces that are much better known. The argument uses a relation between chromo-magnetic matrix elements and twist-three distributions that can be measured in deep inelastic scattering processes. Using (sparse) data on the latter, Ref.~\cite{Seng:2018wwp} found that the unknown direct pieces provide only 10\% corrections to the total matrix element. Neglecting the direct pieces, we find the simple relations that are valid in presence of a Peccei-Quinn mechanism,
\begin{align}
\bar{g}_0(\tilde d_q) &\simeq \delta_{g_0}\frac{1}{4F_{\pi}}\left(\tilde{d}_u+\tilde{d}_d \right)r\frac{d\delta m_N}{d\bar{m}\varepsilon}, \\
\bar{g}_1(\tilde d_q) &\simeq \delta_{g_1}\frac{1}{2F_{\pi}}\left(\tilde{d}_u-\tilde{d}_d \right)r\frac{\sigma_{\pi N}}{\bar{m}},
\end{align}
where we defined the vacuum condensate ratio:
\begin{align}
r=\frac{1}{2}\frac{\langle0|\bar{q}g_s\sigma_{\mu\nu}G^{\mu\nu}q |0\rangle}{\langle0|\bar{q}q |0\rangle},
\end{align}
which has the value $r=(0.4\pm 0.05)~{\rm GeV}^2$ \cite{Ioffe1982,Kogan1984,Narison:2007spa}. Furthermore $(d\delta m_N/d\bar{m}\epsilon)\simeq \delta m_N/(\bar{m}\epsilon)=(2.49\pm0.17\,{\rm MeV})/(\bar m \epsilon)$ \cite{Borsanyi:2013lga,Borsanyi:2014jba}, $\bar{m}=(m_u+m_d)/2=(3.37\pm0.08)~$MeV \cite{Aoki:2016frl} and $\epsilon=(m_d-m_u)/(2\bar{m})=(0.37\pm0.03)$ \cite{Aoki:2016frl}. We have added $\delta_{g_{0,1}}=(1\pm 0.3)$ to account for the theoretical uncertainties in these expressions.

We now turn to  four-quark operators induced, for example, in the mLRSM, cf. Eqs.~(\ref{eq:mHmatching})--(\ref{eq:4quark}). We closely follow Ref.~\cite{Cirigliano:2016yhc} which connected the CP-odd pion-nucleon couplings through chiral symmetry relations to $K\rightarrow \pi \pi$ amplitudes for which lattice-QCD calculations exist. Put together the relevant expressions are given by 
\begin{eqnarray}\label{couplings0}
\bar{g}_1 &=&  \left( 1.45 \pm 0.16\pm 0.75 \right) \times 10^{-5}\,  \textrm{Im} \left(V^*_{us}\frac{ v^2 c^{us}_{Hud}}{\Lambda^2}\right)   +  \left(  2.85 \pm 0.33 \pm 1.5    \right) \times 10^{-5}\,  \textrm{Im} \left(V^*_{ud} \frac{ v^2 c^{ud}_{Hud}}{\Lambda^2}\right)\,,\nonumber\\
\bar{g}_0&=& (0.08 \pm 0.015\pm 0.04)  \times 10^{-5}\, \textrm{Im} \left(V^*_{us}\frac{ v^2 c^{us}_{Hud}}{\Lambda^2}\right)  \,.
\label{couplings1}
\end{eqnarray}
The contributions to the nucleon EDMs are not so well understood and have been calculated with several approaches, see e.g. Refs.~\cite{Zhang:2007da,Maiezza:2014ala}. We follow Ref.~\cite{Cirigliano:2016yhc} and use the next-to-next-to-leading-order expressions for the chiral loops in dimensional regularization \cite{Seng:2014pba}
\begin{eqnarray}
d_n &=& \bar d_n  (\mu) - \frac{e g_A \bar g_1}{8 \pi ^2 F_\pi} \left(  \frac{\bar g_0}{\bar g_1} \left( \log \frac{m^2_\pi}{\mu^2} - \frac{\pi m_\pi}{2 m_N} \right)  + \frac{1}{4 } \left( \kappa_1 - \kappa_0\right) \frac{m^2_\pi}{m_N^2} \log \frac{m^2_\pi}{\mu^2}  \right) \ ,
\label{eq:dn}
\\
d_p &=& \bar d_p (\mu) 
+ \frac{e g_A \bar g_1}{8 \pi^2 F_\pi} \Bigg( \frac{\bar g_0}{\bar g_1} \left( \log \frac{m^2_\pi}{\mu^2} - \frac{2 \pi m_\pi}{m_N} \right)  
-\frac{1}{4 } \left(  \frac{2 \pi m_\pi}{m_N} + \left( \frac{5}{2} + \kappa_1 + \kappa_0\right) \frac{m^2_\pi}{m_N^2} \log \frac{m^2_\pi}{\mu^2}  \right)  \Bigg) 
\ ,\nonumber
\label{eq:dp}
\end{eqnarray}
where $g_A \simeq 1.27$ is the nucleon axial charge, 
and $\kappa_1 = 3.7$ and $\kappa_0 = -0.12$ are related to the nucleon magnetic moments. We set $\mu = m_N$ for the renormalization scale and apply $\bar d_{n,p}   (\mu = m_N) = 0$.  To estimate the uncertainty of this expression we vary the renormalization scale $\mu$ between $m_N$ and $m_K$ in the loop expressions. All expressions for the four-quark operators assume a Peccei-Quinn mechanism and include corrections from the induced $\bar \theta$ term. 

\subsection{The QCD $\bar \theta$ term}
If a pattern of EDMs is observed that can be explained by (small) $\bar\theta$-term alone, it will not be possible to say whether it is the result of an imperfect infrared relaxation of $\bar\theta$ or a small radiative correction to a vanishing ultraviolet boundary condition. This means we can only draw useful conclusions about the strong CP problem if the observable pattern of EDMs is distinguishable from that of a pure $\bar\theta$-term. Thus we must compare EDMs induced in the various beyond-the-SM models discussed above, to those induced by the $\bar \theta$ term alone. 

Let us briefly discuss the hadronic couplings arising from the $\bar \theta$ term. Various lattice-QCD calculations of nucleon EDMs have been reported \cite{Guo:2015tla,Abramczyk:2017oxr,Dragos:2019oxn,Alexandrou:2020mds,Bhattacharya:2021lol}, and the most accurate result was given in Ref.~\cite{Dragos:2019oxn}:
\begin{eqnarray}\label{Latticedn}
d_n &=& -(1.5 \pm 0.7)\cdot 10^{-3} \,\bar \theta\,e\, \mathrm{fm}\,.
\end{eqnarray}
Other recent lattice calculations found EDMs consistent with zero and thus did not confirm these findings \cite{Alexandrou:2020mds,Bhattacharya:2021lol}. Eq.~\eqref{Latticedn} is in agreement with estimates from QCD sum rules and chiral perturbation theory. The proton EDM was calculated on the lattice as well but the result has a larger uncertainty. Instead we apply a relation $d_p = -(1\pm 0.5) d_n$ which covers various estimates from the chiral perturbation theory, QCD sum rules, and the lattice results. 

The CP-odd pion-nucleon couplings are better under control. Ref.~\cite{deVries:2015una} found
\begin{eqnarray}\label{g0theta}
\bar g_0 &=& -(14.7\pm2.3)\cdot 10^{-3}\,\bar \theta\,,\nonumber\\
 \bar g_1 &=&\phantom{-} (3.4\pm2.4)\cdot 10^{-3}\,\bar \theta\,,
\end{eqnarray}
through chiral symmetry relations between CP-odd pion-nucleon interactions and quark-mass corrections to baryon masses.

Finally, Ref.~\cite{Flambaum:2019ejc} noticed that a nonzero $\bar \theta$ term induced CP-odd electron-nucleon interactions through electromagnetic loops. They obtained \begin{equation}
C_S = -(3\pm1.5)\cdot 10^{-2}\,\bar \theta
\end{equation} where $C_S$ is a linear combination of $C_S^{(0)}$ and $C_S^{(1)}$ appearing in Eq.~\eqref{hadronicCPV}. We will discuss this combination in more detail in the next section. 

\subsection{EDMs of nuclei, atoms, and molecules}
We are now in the position to calculate EDMs of various systems. The nucleon EDMs have already been discussed above so we turn to larger systems. We begin with paramagnetic systems that are mainly sensitive to (semi-)leptonic CP-violating operators. The most stringent limits are from polar molecules due to huge inner-molecular electric fields induced by relatively small external electric fields. A nonzero electron EDM and/or electron-nucleon interactions affect the frequency associated with the response of paramagnetic polar molecules to such an applied external field. The frequency $\omega$ is given by
\begin{align}
\omega&=\alpha_{d_e}d_e+\alpha_{C_S}C_S,
\end{align}
where the coefficient $C_S$ is given by 
\begin{align}
C_S\equiv C^{(0)}_S+\frac{Z-N}{Z+N}C^{(1)}_S.
\end{align}
Here, $Z$ and $N$ are the proton and neutron numbers of the heaviest atom of the molecule. The parameters $\alpha_{d_e}$ and $\alpha_{C_S}$ depend on the paramagnetic molecular system of interest. As an example, Table \ref{alpha_deCs} presents the values of $\alpha_{d_e}$ and $\alpha_{C_S}$ in ThO, HfF$^+$ and BaF systems
\cite{doi:10.1063/1.4968597,Skripnikov,PhysRevA.96.040502,doi:10.1063/1.4968229,PhysRevA.90.022501,PhysRevA.93.042507,Malika:2019jhn,haase2021systematic}.
\renewcommand{\arraystretch}{1.5}
\begin{table}[t!]
\centering
\small
\begin{tabular}{|c c c|}
\hline
 & $\alpha_{d_e}$  & $\alpha_{C_S}$  \\
\hline
ThO & \raisebox{1pt}{$\frac{(120.6\pm 4.9)~{\rm mrad}/{\rm s}}{10^{-27}~e~{\rm cm}}$} & \raisebox{1pt}{$(181.6\pm 7.3)\times10^{7}~{\rm mrad/s}$}  \\
HfF$^+$ & \raisebox{2pt}{$\frac{(34.9\pm 1.4)~{\rm mrad/s}}{10^{-27}~e~{\rm cm}}$}  & \raisebox{2pt}{$(32.0\pm 1.3)\times10^7~{\rm mrad}/{\rm s}$}  \\
BaF & \raisebox{2pt}{$\frac{(19.7\pm 0.75)~{\rm mrad/s}}{10^{-27}~e~{\rm cm}}$}  & \raisebox{2pt}{$(12.7\pm 0.18)\times10^7~{\rm mrad}/{\rm s}$} \\
\hline
\end{tabular}
\caption{Input parameters for EDMs of paramagnetic molecules. }
\label{alpha_deCs}
\end{table}

EDMs of nuclei are only sensitive to hadronic sources of CP violation. So far no nuclear EDMs have been measured, but there are plans to measure the EDMs of light nuclei in storage rings \cite{CPEDM:2019nwp}. Anticipating such measurements we consider the deuteron EDM \cite{Bsaisou:2014zwa,Yamanaka:2015qfa}
\begin{eqnarray}
d_{D} &=&
(0.94\pm0.01)(d_n + d_p) + \bigl [ (0.18 \pm 0.02) \,\bar g_1\bigr] \,e \,{\rm fm} \, .
 \label{eq:h2edm} 
\end{eqnarray}
Diamagnetic atomic EDMs are sensitive to nuclear CP violation and electron-nucleon interactions.  ${}^{225}$Ra, due to its octopole deformation, is mainly sensitive to CP-odd nuclear forces that are, in turn, dominated by one-pion-exchange processes involving $\bar g_{0,1}$. We use 
\cite{Engel:2013lsa,Dobaczewski:2018nim}
\begin{eqnarray}\label{dHg}
d_{\mathrm{Ra}} &=& (7.7\cdot 10^{-4})\cdot\left[(2.5\pm 7.5)\,\bar g_0 - (65 \pm 40)\,\bar g_1\right]e\, {\rm fm}.
\end{eqnarray}
The situation is more complicated for the diamagnetic atom ${}^{199}$Hg. This system gets relevant contributions from the nucleon EDMs, the CP-violating nuclear force, the electron EDM, and from CP-odd electron-nucleon interactions. We write \cite{Dmitriev:2003kb,Engel:2013lsa,Yamanaka:2017mef,Fleig:2018bsf,2018arXiv180107045S}
\bea\label{dHg} d_{\rm Hg}&=& (0.012\pm0.012) d_e - \left[ (0.028\pm0.006) C_S- \frac{1}{3}(3.6\pm0.4) \left(C_T+\frac{Z\al}{5 m_N R}C_P\right)\right]\cdot
10^{-20}\, e\,\mathrm{cm}\nonumber\\
&&  -(2.1\pm0.5)
\Ex{-4}\bigg[(1.9\pm0.1)d_n +(0.20\pm 0.06)d_p+\bigg(0.13^{+0.5}_{-0.07}\,\bar g_0 +
0.25^{+0.89}_{-0.63}\,\bar g_1\bigg)e\, {\rm fm}\bigg]\nn\\
\eea
where  $R\simeq 1.2\, A^{1/3}$ fm is the nuclear radius, and 
$C_{P,T} = (C_{P,T}^{(n)}\langle \vec \sigma_n\rangle  +C_{P,T}^{(p)}\langle \vec \sigma_p\rangle
)/(\langle \vec \sigma_n\rangle +\langle \vec \sigma_p\rangle )$, with $C_{P,T}^{(n,p)}=C_{P,T}^{(0)}\mp C_{P,T}^{(1)}$. For $^{199}$Hg
we have \cite{Yanase:2018qqq} 
\bea
\langle \vec \sigma_n\rangle= -0.3249\pm0.0515\,,\qquad \langle \vec \sigma_p\rangle = 0.0031\pm 0.0118\,.
\eea

\section{Disentangling sources of CP violation}\label{results}
In this section we discuss how EDM measurements of various systems can be used to separate various CP-violating BSM models and the SMEFT operators they generate from a pure $\bar \theta$ scenario.  The former, we have argued, require an infrared relaxation of $\bar\theta$. The latter, however, is not necessarily indicative of a PQ mechanism, since it could be the remnant of an imperfect UV solution, for instance due to spontaneous breaking of P or CP. In addition, as we will discuss, we also have to distinguish scenarios where CP violation is dominated by dimension-six operators that are not necessarily correlated with large threshold corrections to $\bar \theta$. In the models we study the prime example is an electron EDM, which does not appear in Eq.~\eqref{eq:thetacorr}.

\subsection{Case 1:  Scalar Leptoquark Models}
We begin with the simple LQ scenario discussed in Sect.~\ref{sec:LQ} and focus on couplings between first-generation quarks and leptons. That is, we set $x^{eu}_{LR}=x^{u e}_{RL}= e^{i \alpha}$ where $\alpha$ is some nonzero phase, and set the remaining couplings to zero. In this simple case, the dominant CP-violating dimension-six operators are quark-electron interactions, while the electron EDM is one-loop suppressed, which is not compensated by the slight enhancement by $m_u/m_e$. The quark EDMs and chromo-EDMs also appear only at one loop and are further suppressed by a factor of $m_e/m_u$. 

\begin{figure}[t!]
\begin{center}
\includegraphics[scale=0.5]{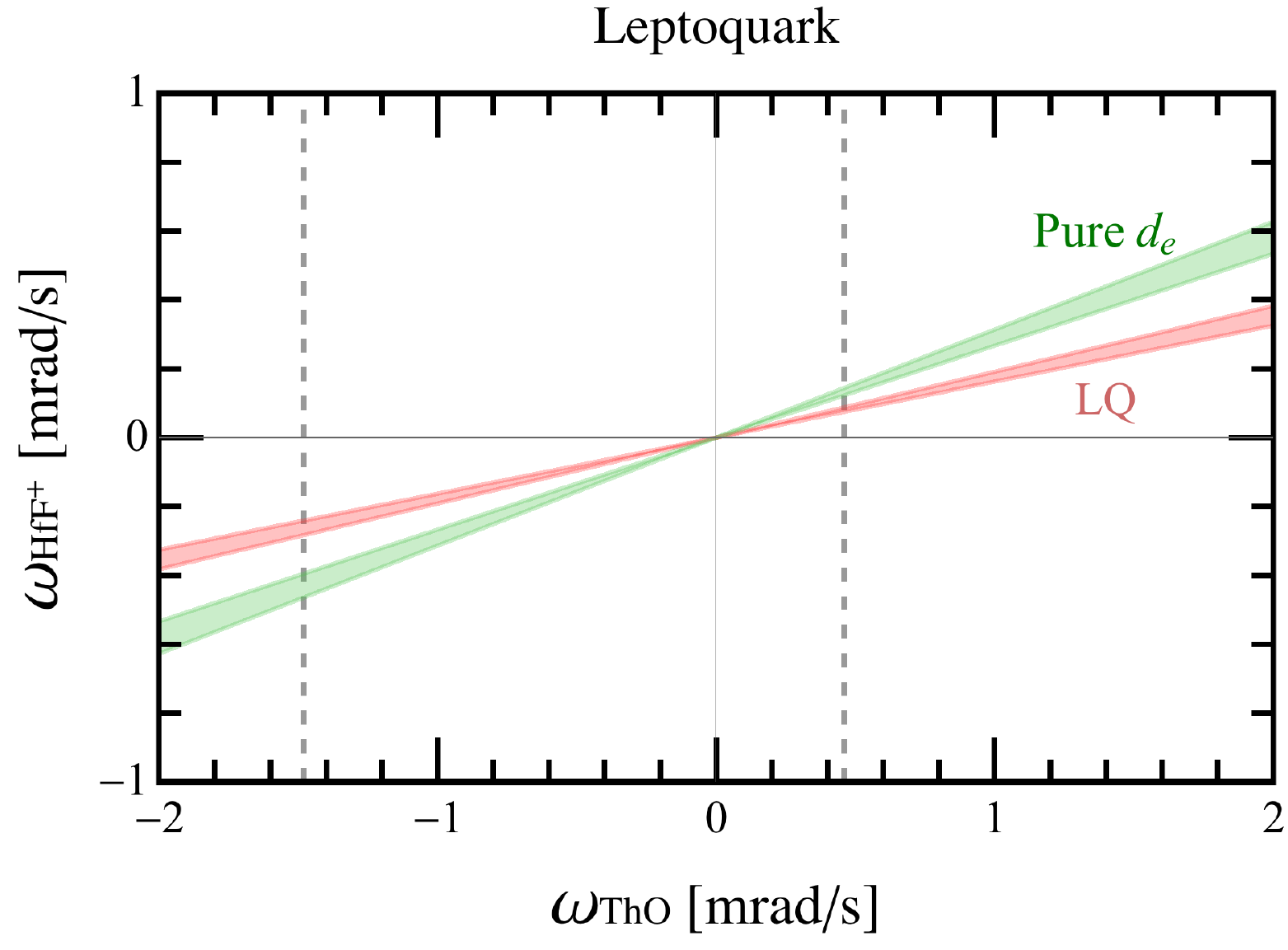}
\includegraphics[scale=0.51]{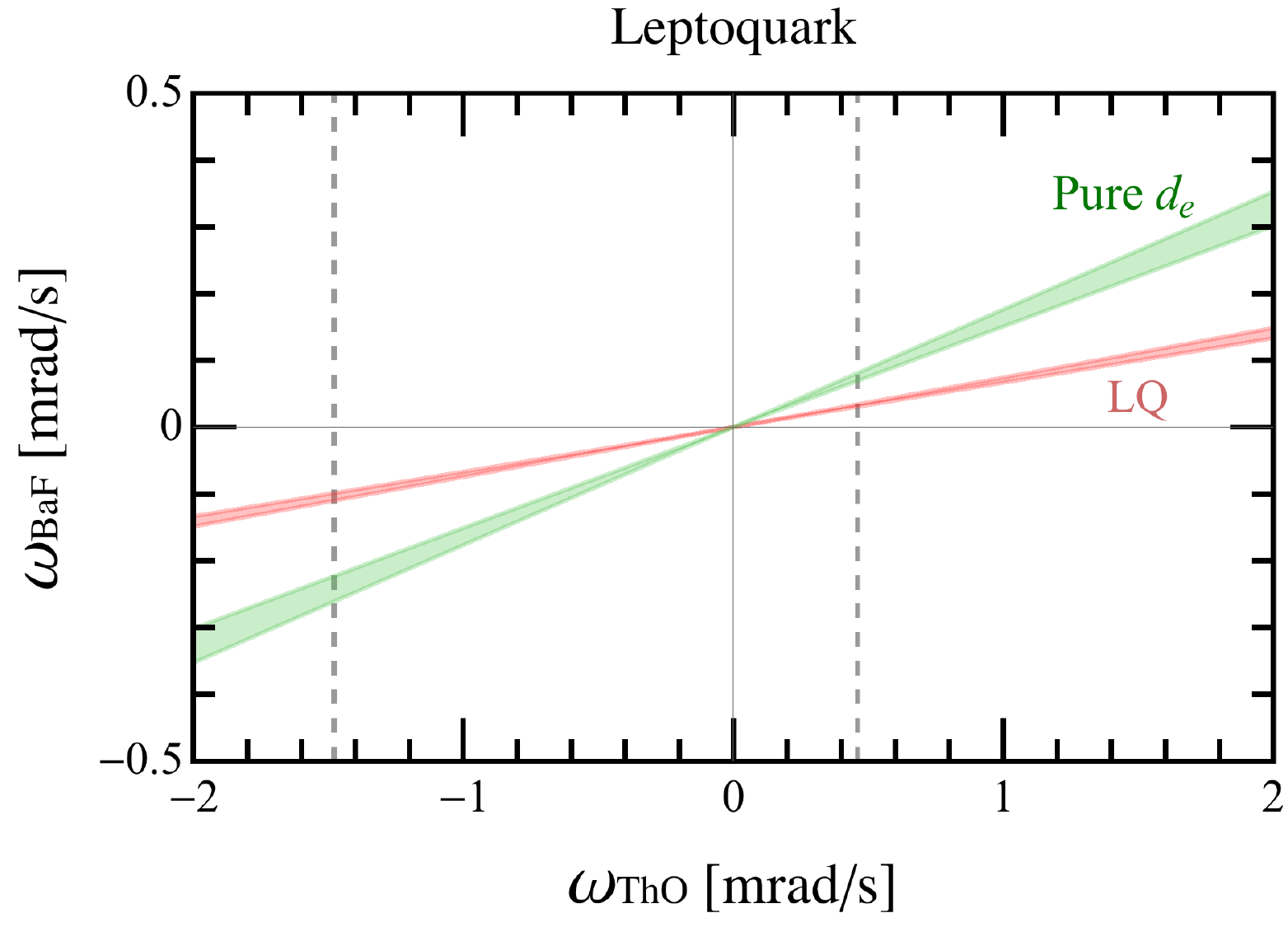}
\end{center}
\caption{The left (right) panel shows the correlation between $\omega_{\rm HfF^+}$ ($\omega_{\rm BaF}$) and $\omega_{\rm ThO}$ in a pure-$d_e$ scenario (green) and the LQ scenario (red). The bands indicate the small uncertainty arising from hadronic and molecular matrix elements. The region between the vertical dashed lines is allowed at $2\sigma$ level.}
\label{dpara_LQ}
\end{figure}

This scenario precisely predicts  ratios between EDMs of paramagnetic systems \cite{Chupp:2014gka,Fleig:2018bsf}. This is illustrated in Fig.~\ref{dpara_LQ} where the red bands in the left and right panels illustrate the ratios $\omega_{\rm HfF^+}/\omega_{\rm ThO}$ and $\omega_{\rm BaF}/\omega_{\rm ThO}$, respectively. The error bands are small as both the hadronic matrix elements connecting electron-quark to electron-nucleon couplings as well as the molecular theory is well under control. The contributions from the electron EDM are very small and appear at the $10^{-4}$ level. Therefore, we can apply the present situation to a sole-source limit in which the bound on $C_S$ is obtained by assuming $d_e=0$. Taking the current limit, $|C_S|<7.3\times 10^{-10}$ ($90\%$ C.L.) \cite{Andreev:2018ayy}, we obtain $m_{R_2}\gtrsim 1.9\times 10^4~$TeV when $|x_{RL}^{eu}x_{RL}^{ue}|\sin\alpha=1$, illustrating the excellent sensitivity of EDM experiments to new sources of CP violation.

It is possible to separate this scenario from one where the dominant source of CP violation is the electron EDM. This is, for instance, the case in a LQ model where $x^{et}_{LR}=x^{t e}_{RL}= e^{i \alpha}$ with the other couplings set to zero. In this case the electron EDM in Eq.~\eqref{oneloopLQ} is enhanced by $m_t/m_e$ and provides the dominant contribution to paramagnetic systems. The induced electron-gluon operators provide negligible contributions. 
In this case, the renormalization of $\bar \theta$ is not excessive, as expected from the EFT arguments of Ref.~\cite{deVries:2018mgf}, and thus a $d_e$-dominated EDM pattern would not necessarily point towards an IR solution of the strong CP problem. A $d_e$-dominated scenario like this leads to the green bands in Fig.~\ref{dpara_LQ}. Clearly, sufficiently precise measurements of two paramagnetic systems can separate these scenarios. The current upper limit on $\omega_{\rm HfF^+}$ \cite{Cairncross:2017fip} is a few times weaker than that of $\omega_{\rm ThO}$ \cite{Andreev:2018ayy}, but in general if a nonzero signal is found in one system, the other one should be around the corner. No measurements of  $\omega_{\rm BaF}$ exist, but an experiment is in development \cite{Aggarwal:2018pru}. 

\begin{figure}[t!]
\begin{center}
\includegraphics[scale=0.5]{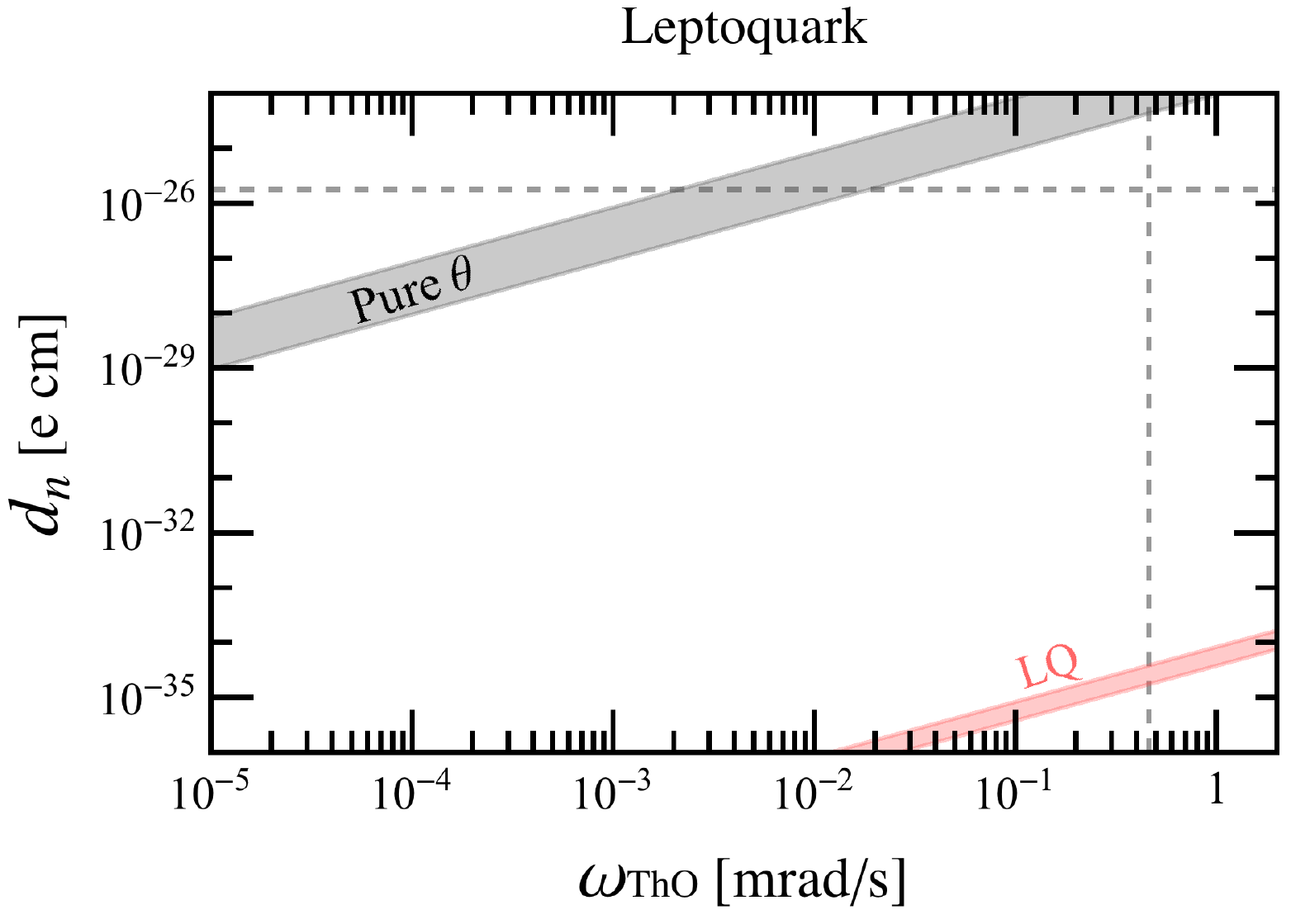}
\end{center}
\caption{Correlation between $\omega_{\rm ThO}$ and $d_n$ in the LQ model (red) and a pure-$\bar \theta$ scenario (gray). The dashed lines correspond to experimental bounds.}
\label{puretheta_LQ}
\end{figure}

With just paramagnetic systems it is not possible to separate a pure-$\bar \theta$ scenario from the LQ scenario with couplings to first-generation quarks. In both cases, the paramagnetic observables are dominated by the $C_S$ interactions, and the ratios are therefore almost identical. However, in this case the neutron and diamagnetic EDMs provide a way out. In a pure-$\bar \theta$ scenario such EDMs are relatively large compared to paramagnetic EDMs, whereas in the LQ scenario hadronic CP violation is loop suppressed. We illustrate this in Fig.~\ref{puretheta_LQ} where we depict $\omega_{\rm ThO}$ versus $d_n$. The bands for the LQ and pure-$\bar \theta$ are very distinct. For instance, if a next-generation ThO measurement finds a nonzero value $\omega_{\rm ThO} = 0.1$ mrad/s, then a pure $\bar \theta$ scenario is already excluded as this would predict $d_n \simeq 3 \cdot 10^{-25}$ e cm, well above the existing limit $d_n \leq 1.8 \cdot 10^{-26}$ e cm \cite{Abel:2020gbr}, depicted by the horizontal dashed line. For Hg similar statements hold but here the hadronic and nuclear uncertainties make definite statement more complicated.

In summary, the simple $R_2$ leptoquark model illustrates the EFT arguments of~\cite{deVries:2018mgf}. A pattern of EDMs indicative of higher-dimensional CP-odd operators that mix quadratically with $\bar \theta$ is directly associated with large threshold corrections to $\bar \theta$ at the EFT matching scale. The large threshold corrections must be relaxed in the infrared. The leptoquark model also provides an example where paramagnetic EDMs are dominated by a dimension-six operator, the electron EDM, that does not indicate  large corrections to $\bar \theta$. However, such scenarios can be separated through precise measurements of at least two paramagnetic EDMs.

\subsection{Case II:  The MSSM}

Since we have focused on a limiting case of the MSSM where the only phase is in the gluino mass, 
the EDM phenomenology is dominated by quark EDMs and chromo-EDMs. This implies that paramagnetic systems are not particularly relevant for the discussion and we can focus on nucleon, nuclear, and diamagnetic systems. While we have expressed our results in terms of several MSSM parameters, in all ratios of EDMs the dependence on $m_{\rm soft}$ and $\phi_3$ cancels and the only dependence is on the squark mass splitting parametrized by $\Delta$. This dependence is actually mild for $\Delta <1$, where the functions $F(\Delta)$ and $G(\Delta)$ only vary by about $35\%$, and the ratio of functions $F(\Delta)/G(\Delta)$ only varies by a factor of 2 for any $\Delta$. We assume that an ${\cal{O}}(1)$ splitting among the squarks is a reasonably general region of parameter space and we take $\Delta =1$ for concreteness.
Before we discuss ratios of EDMs we briefly mention the reach of present EDM experiments. The current $d_{\rm Hg}$ limit results in $m_{\rm soft}\gtrsim 20~$TeV at $\phi_3=\pi/2$, taking central values for the hadronic and nuclear matrix elements. The experimental bound on $d_n$ leads to the somewhat weaker limit $m_{\rm soft}\gtrsim 5.4~$TeV but has smaller uncertainties.

The neutron and proton EDMs are dominated by the quark chromo-EDMs while the quark EDMs provide roughly 20\% corrections. The ratio of the neutron- and proton-induced chromo-EDM contributions depends on the ratio
\begin{equation}
\frac{d_p(\tilde d_q)}{d_n(\tilde d_q)}=\tilde c_p \frac{\left(4 Q_d \tilde d_d - Q_u \tilde d_u\right)}{\left(- 4 Q_d \tilde d_u + Q_u \tilde d_d\right)} \simeq -0.78\cdot(1\pm 0.2)\,,
\end{equation}
where $\tilde c_p = 1\pm0.2$ capture the uncertainty due to isospin breaking. Unfortunately, for $\bar \theta$-dominated EDMs,
\begin{equation}
\frac{d_p(\bar \theta)}{d_n(\bar \theta)}=-1\cdot(1\pm 0.5)\,.
\end{equation}
Thus these ratios overlap within uncertainties, implying that measurements of both nucleon EDMs cannot separate a pure $\bar \theta$ scenario from our MSSM model. This is clearly illustrated in the top-left panel of Fig.~\ref{MSSMplots2}, where also qEDM contributions are included, where it can be seen that the gray and red bands overlap entirely.

Nuclear and diamagnetic systems are much more promising, as in these cases the contribution from the CP-odd nuclear force can break the degeneracy. For instance, in the deuteron EDM the contributions from $d_n + d_p$ are small in both the MSSM and the pure-$\bar \theta$ scenario. However, the contribution from CP-odd one-pion-exchange proportional to $\bar g_1$ is rather different in both models. $\bar g_1$ is relatively small for $\bar \theta$, since the $\bar\theta$ term is an isospin-conserving interaction and the generation of $\bar g_1$ is suppressed by the small quark mass difference over the chiral-symmetry-breaking scale $\Lambda_\chi \sim 1$ GeV. In the MSSM however, the isospin-breaking combination $\tilde d_u-\tilde d_d$ induces $\bar g_1$ directly.  As such $|d_D/d_n| \gtrsim O(1)$ for the MSSM \cite{Lebedev:2004va,Dekens:2014jka} whereas $|d_D/d_n| \lesssim O(1)$ for $\bar \theta$. This is reflected in the top-right panel of Fig.~\ref{MSSMplots2} where, despite sizable hadronic uncertainties, the gray and red bands do not overlap.

A deuteron EDM measurement would be ideal for our goals as it is theoretically relatively clean. However, there is no competitive measurement planned on short time scales, although plans exist for storage-ring experiments. We therefore consider EDMs of diamagnetic atoms. We begin with ${}^{225}$Ra, which is analogous to the deuteron EDM as it is dominated by contributions from the isospin-breaking CP-violating nuclear force proportional to $\bar g_1$. Unfortunately, unlike for the deuteron EDM, the nuclear uncertainties are sizable. This is reflected in the bottom-left panel of Fig.~\ref{MSSMplots2} where the red and gray bands are wider than was the case for the deuteron. Still, they may be separable. On the other hand, an upper bound on the ${}^{225}$Ra EDM has already been set: $d_{\rm Ra} < 1.2\cdot 10^{-23}\,e\,\text{cm}$, and a measurement with sensitivity at the $10^{-28}\,e\,\text{cm}$ level is foreseen \cite{Bishof:2016uqx}. If such an experiment sees a signal, the MSSM scenario would predict a somewhat larger neutron EDM than in the pure $\bar \theta$ scenario. Slight improvements of nuclear matrix elements would reduce the width of the MSSM bands and make the statements more definite. 

Competitive measurements of $d_D$ and $d_{\rm Ra}$ are still in the future, and the most stringent limit is that on $|d_{\rm Hg}| \leq 6.3 \cdot 10^{-30}\,e\,\text{cm}$ \cite{Graner:2016ses}. Unfortunately, the nuclear uncertainties are severe and make it hard to separate the MSSM from the pure-$\bar \theta$ scenario using $d_{\rm Hg}$. This is illustrated in the bottom-right panel of Fig.~\ref{MSSMplots2} where the red and gray bands overlap. Improved calculations of the nuclear matrix elements would go along way in making the predictions more robust. In any case, despite large uncertainties, the pure-$\bar \theta$ scenarios can be excluded if measurements fall outside the gray bands. Such a result would be sufficient to make the point.

 \begin{figure}[t!]\begin{center}
 \includegraphics[scale=0.5]{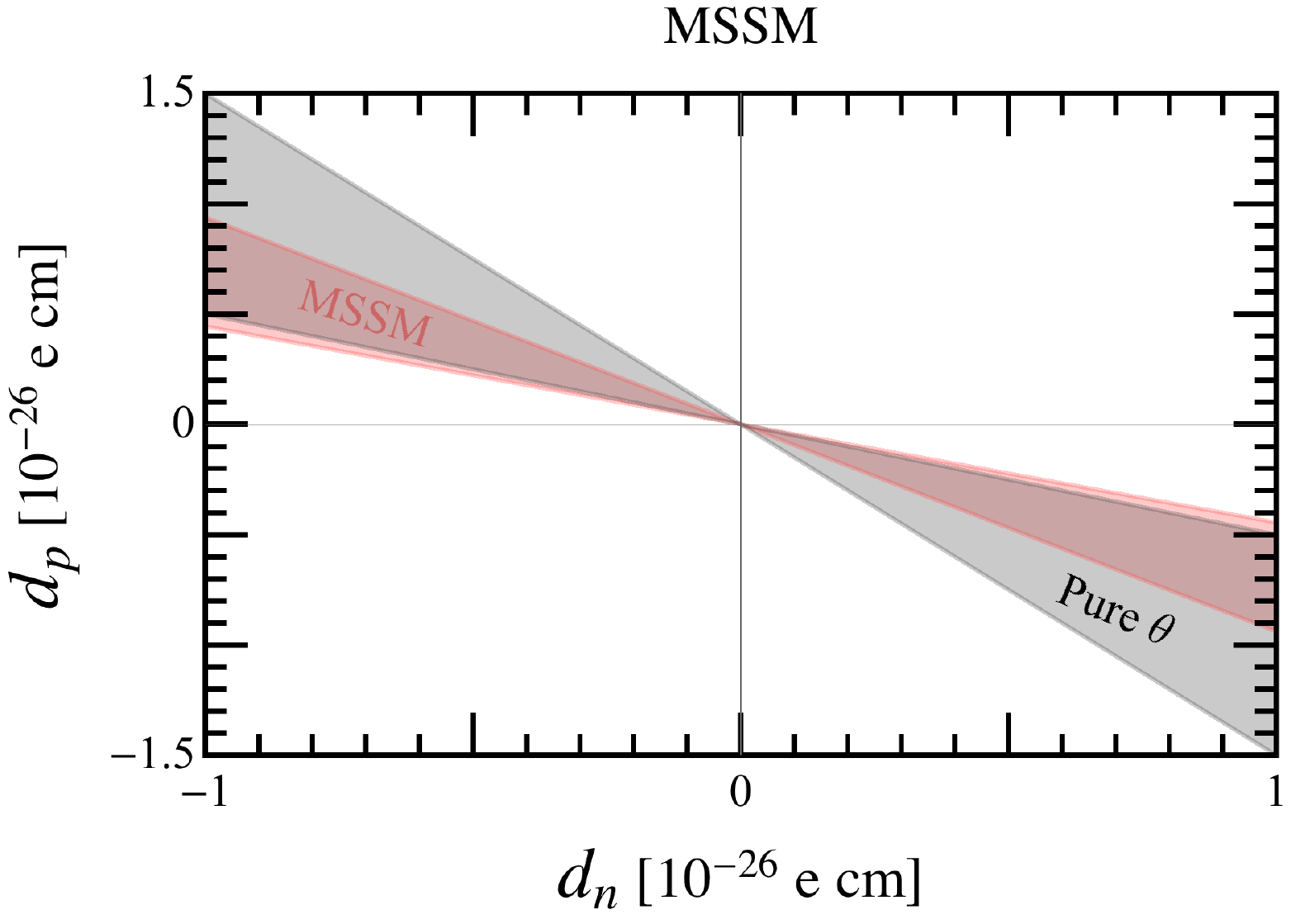}
\includegraphics[scale=0.5]{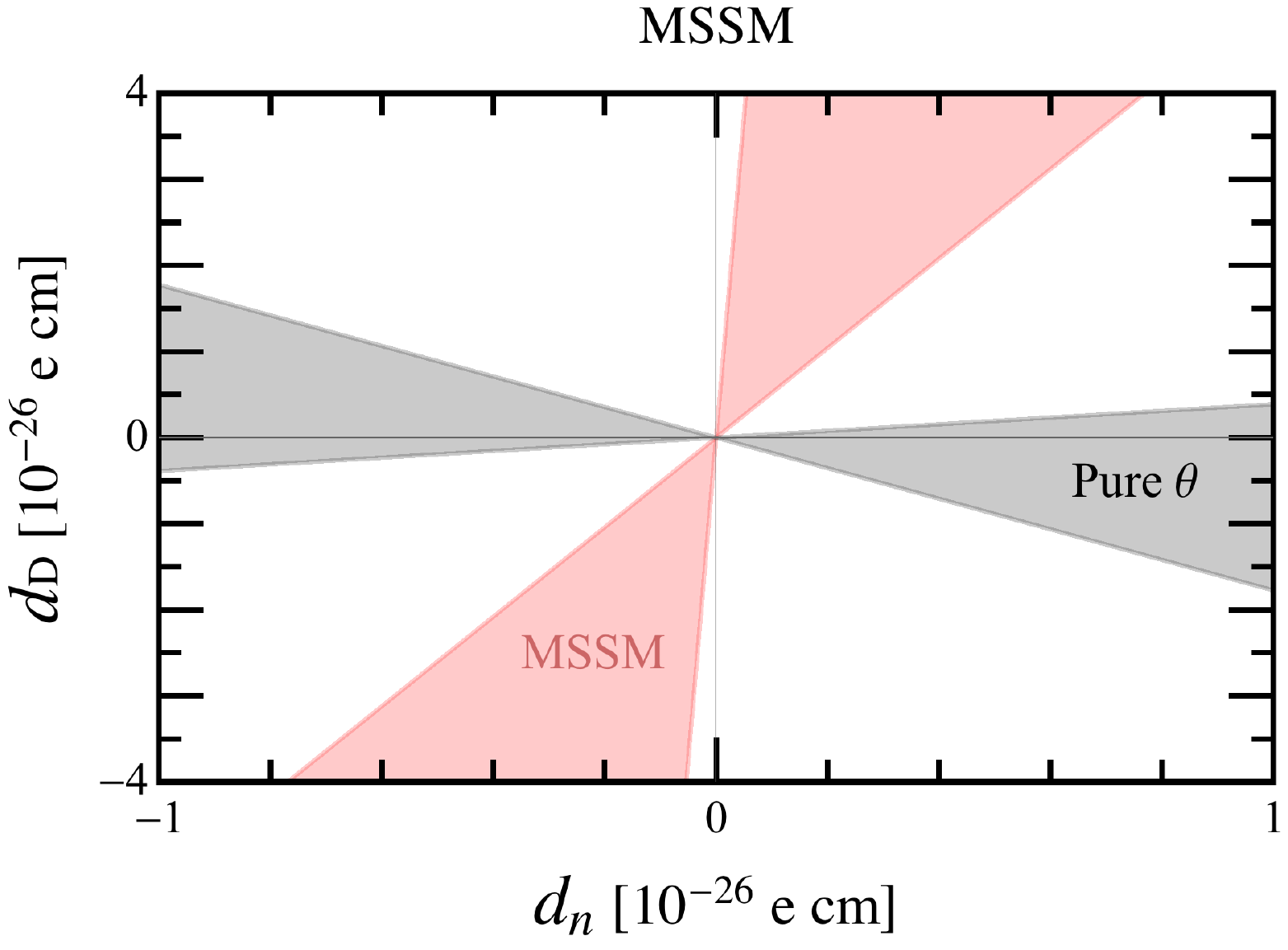}
\includegraphics[scale=0.5]{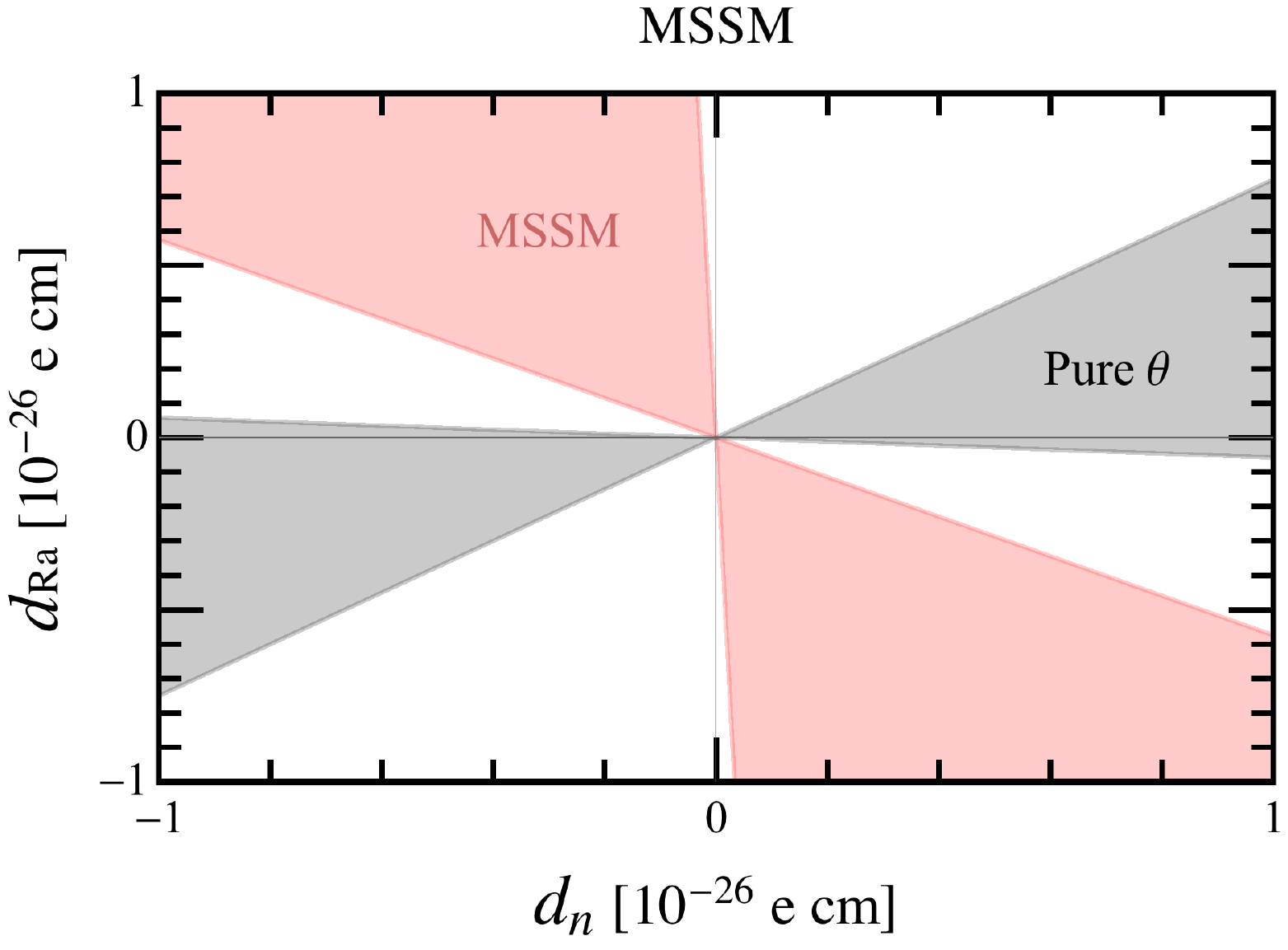}
\includegraphics[scale=0.5]{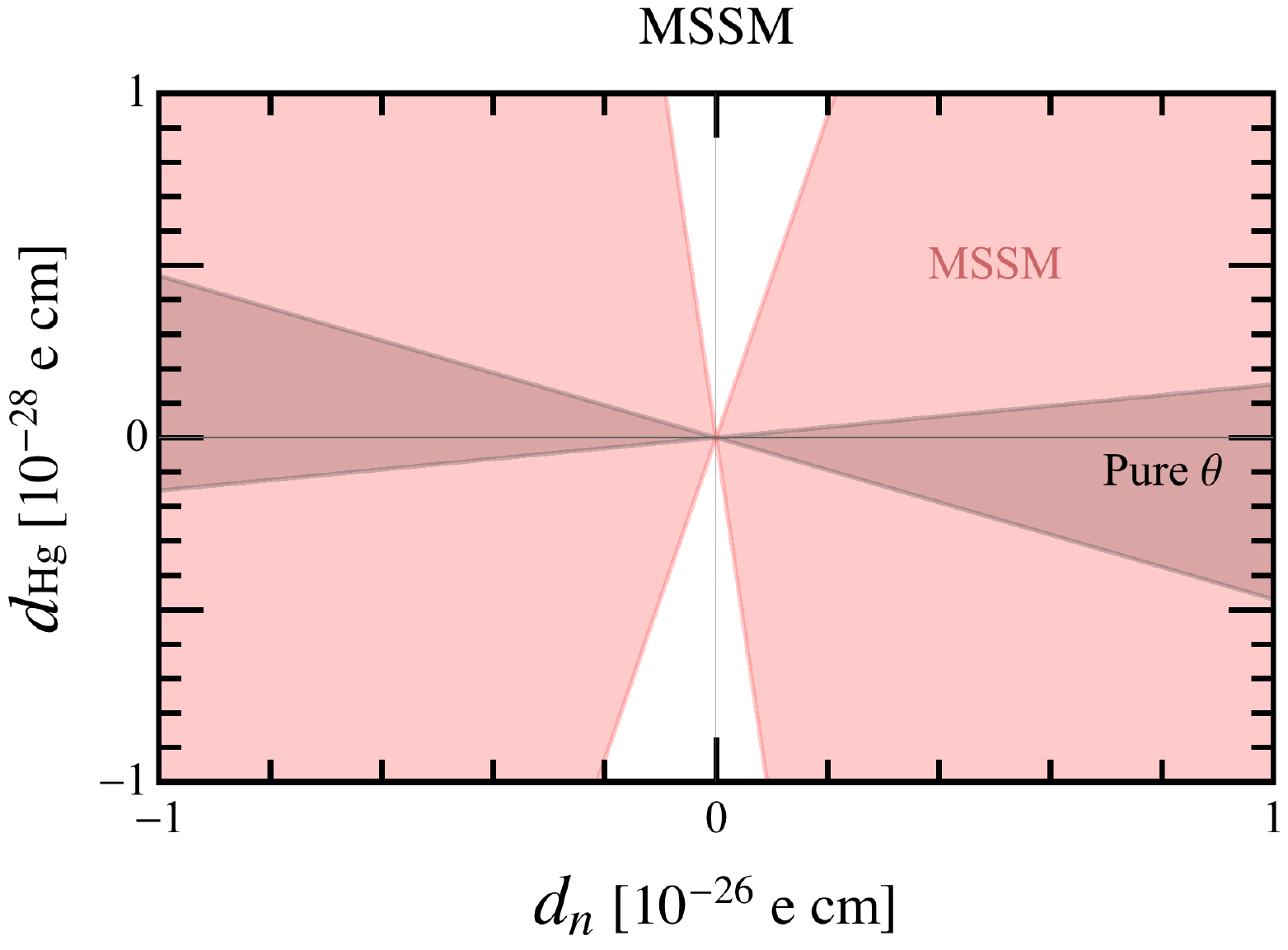}
\end{center}
\caption{Correlation between various EDMs and the neutron EDM for a pure-$\bar \theta$ scenario (gray) and the MSSM (red). The bands indicate the uncertainty in the ratios arising from hadronic and nuclear matrix elements .}
\label{MSSMplots2}
\end{figure}

\subsection{Case III:  The P-symmetric minimal left-right-symmetric model}

In the mLRSM the dominant contribution to EDMs arises at tree-level from the CP-violating four-quark operators in Eq.~\eqref{eq:4quark}. While most literature focuses on the neutron EDM \cite{Zhang:2007da,Maiezza:2014ala,Seng:2014pba}, nuclear and diamagnetic atoms are actually larger due to the large contribution to $\bar g_{1}$ and to lesser extent $\bar g_0$ in Eq.~\eqref{couplings0}. It was already pointed out in Refs.~\cite{deVries:2012ab, Dekens:2014jka} that this leads to enhanced nuclear EDMs, e.g. $|d_D| \gg |d_n|$, and that several measurements can separate contributions from $\bar \theta$ and the four-quark operators \cite{Dekens:2014jka,Ramsey-Musolf:2020ndm}. 
This can already be glimpsed from the different sensitivities of  $d_n$ and $d_{\rm Hg}$ to the left-right symmetry-breaking scale. We set\footnote{We require $\xi \sin \alpha \leq m_b/m_t$ to account for the observed quark masses \cite{Maiezza:2010ic}.} $\xi \sin \alpha \simeq m_b/m_t$, and using central values of the hadronic and nuclear matrix elements obtain  $m_{W_R}\gtrsim 126~(18)~$TeV from $d_{\rm Hg}$ ($d_n$). 

We further illustrate this point in Fig.~\ref{LRmodel}. The $d_p/d_n$ ratio is not well predicted in the mLRSM due to large uncertainties from short-distance contributions not captured by pion loops. In addition, the mLRSM and pure-$\bar \theta$ ratio bands for $d_p/d_n$ overlap within uncertainties. 

The story is completely different for larger systems. The four-quark operators induced in the mLRSM lead to a sizable CP-odd nuclear force arising from one-pion-exchange diagrams proportional to $\bar g_1$. Both the hadronic and nuclear matrix elements are well under control leading to a ratio $d_D/d_n = \mathcal O(50)$, which is very different from the QCD $\bar \theta$ term that predicts much smaller values for the same ratio. The same holds for the diamagnetic atom ${}^{225}$Ra, but in this case the nuclear matrix elements are more uncertain leading to a broader band. Nevertheless, the ratio $d_{\rm Ra}/d_n$ are very distinct in the mLRSM with respect to the pure-$\bar \theta$ scenario. In principle, this would hold for $d_{\rm Hg}/d_n$ as well were it not for the large nuclear uncertainties. As a result, the mLRSM and $\bar \theta$ bands overlap for their $d_{\rm Hg}/d_n$ predictions. 

Finally, let us discuss how a UV solution to the strong CP problem would work in the $P$-symmetric mLRSM. The tree-level correction to $\bar \theta$ can be made small by picking a sufficiently small value for the spontaneous phase $\alpha$. In essence this simply transfers the question of the origin of the smallness of $\bar \theta$ in the SM, to that of the smallness of $\alpha$. Nevertheless, some authors have argue this should be considered as a type of solution \cite{Senjanovic:2020int}. In such a UV solution, the CP-odd part of the dimension-six operator in Eq.~\eqref{dim6edms} scales with $\alpha$ as well and is suppressed by $v^2/v_R^2$ compared to the tree-level $\bar \theta$ term. As such, the low-energy EDM phenomenology is dominated by $\bar \theta$ \cite{Maiezza:2014ala}. This is consistent with our assertion that only the presence of dimension-six contributions to EDMs arising from operators in Eq.~\eqref{eq:thetacorr} can be used to infer that strong CP is solved by infrared relaxation.
\begin{figure}[t!]
\begin{center}
\includegraphics[scale=0.5]{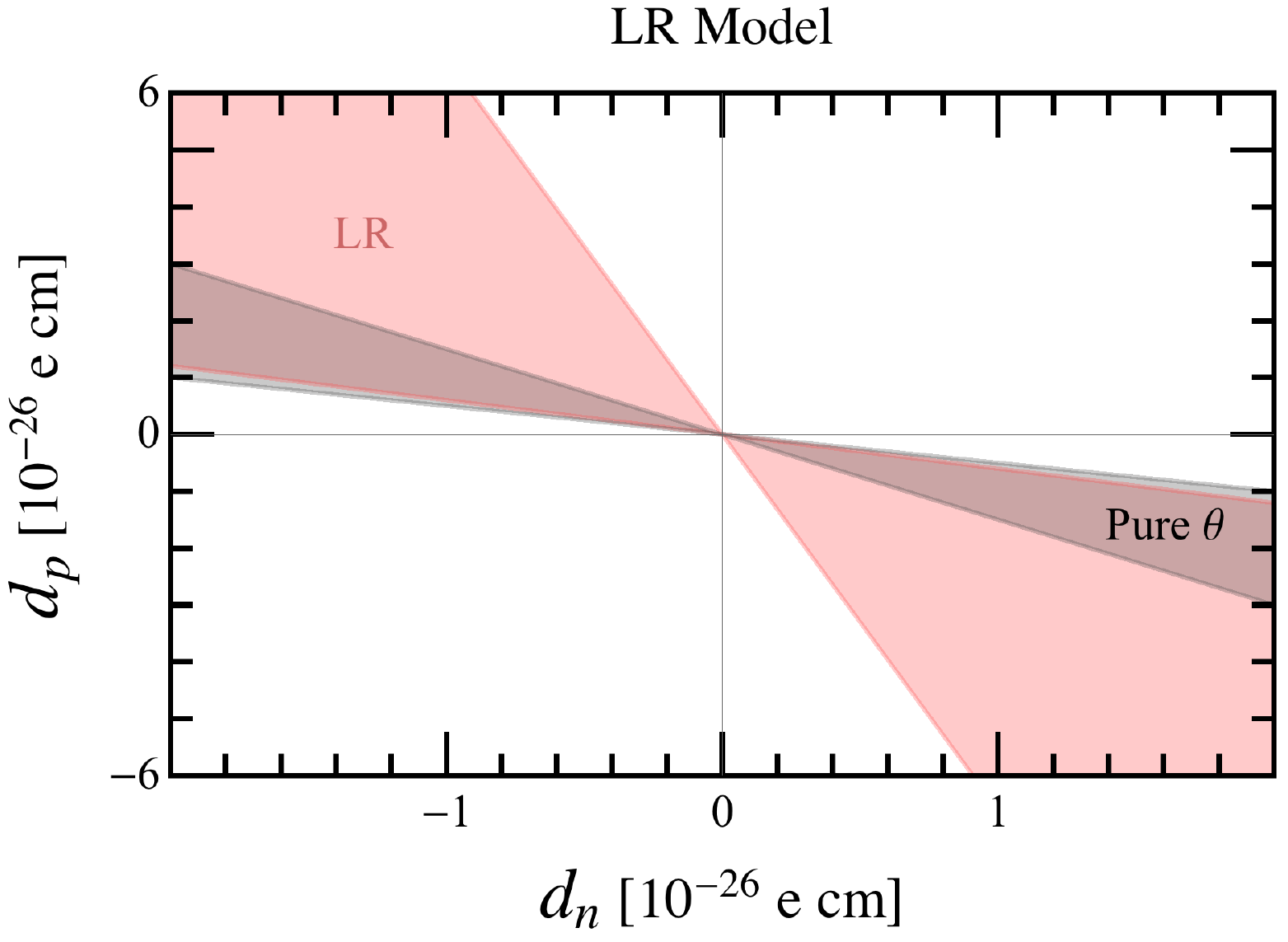}
\includegraphics[scale=0.5]{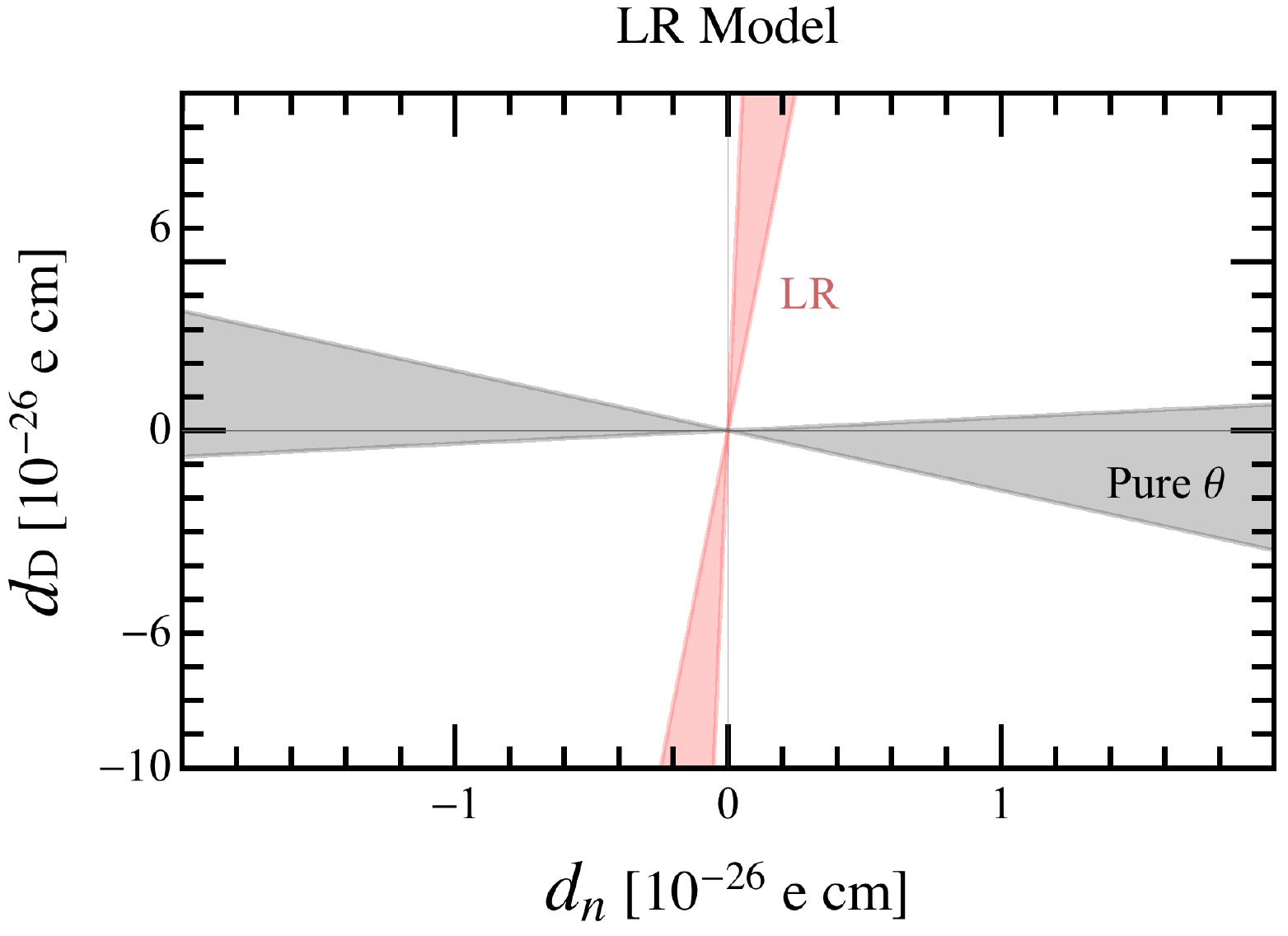}
\includegraphics[scale=0.5]{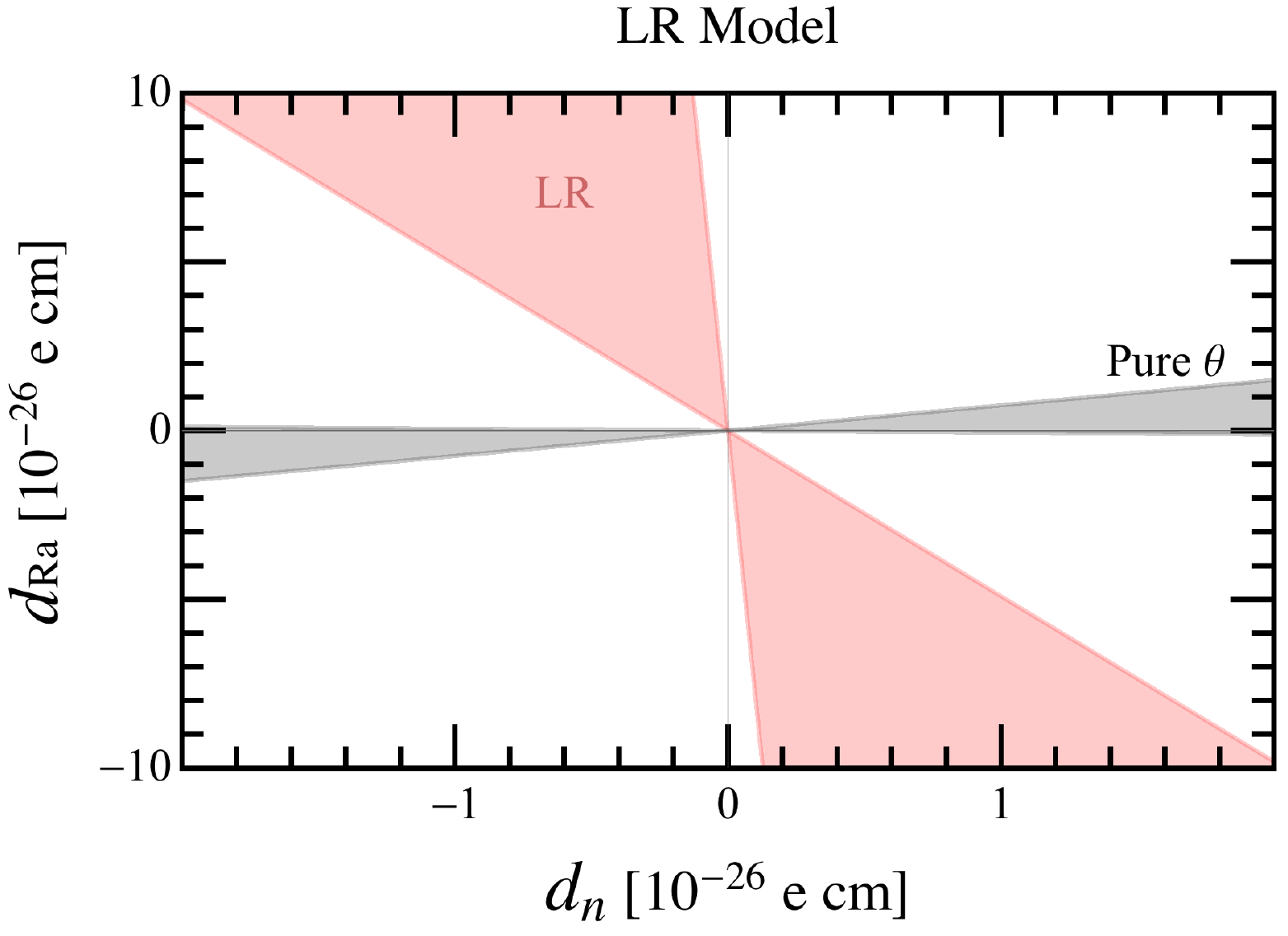}
\includegraphics[scale=0.5]{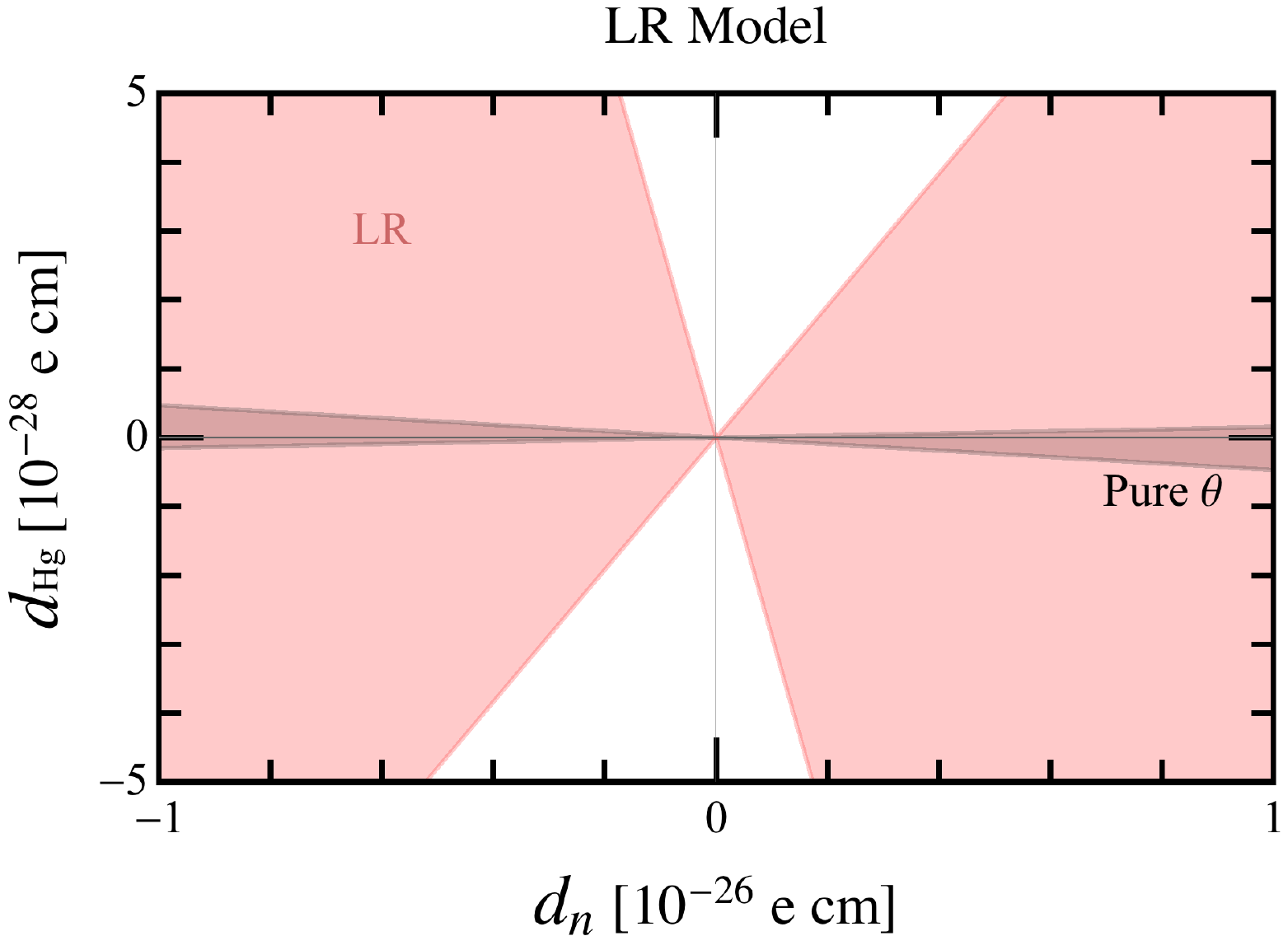}
\end{center}
\caption{Correlation between various EDMs and the neutron EDM for a pure-$\bar \theta$ scenario (gray) and the mLRSM (red). The bands indicate the uncertainty in the ratios arising from hadronic and nuclear matrix elements. }
\label{LRmodel}
\end{figure}

\section{Conclusions}\label{conclusions}
The strong CP problem has been getting worse by about an order of magnitude every decade for the last sixty years~\cite{Smith:1957ht}. The main experimental probes are the neutron  and ${}^{199}\rm{Hg}$ EDMs, which establish the severity of the problem, and searches for axion dark matter. In this work, building on Ref.~\cite{deVries:2018mgf}, we have attempted to strengthen the argument that a diverse portfolio of hadronic EDM measurements could provide valuable data on the mechanism that addresses strong CP. The basic observation is that it is unnatural to have both new, observably large  sources of hadronic CP violation, and an ultraviolet solution to strong CP. If a pattern of hadronic EDMs is observed that is inconsistent with a small value of $\bar\theta$ alone, infrared relaxation by an axion becomes the most natural solution. At the same time, if a Peccei-Quinn mechanism addresses strong CP, there is no particular reason for BSM physics to preserve CP to high accuracy. 

We have surveyed a handful of BSM models to illustrate this effective field theory argument in detail: when there are large phases, there is a large threshold correction to $\bar\theta$ that requires infrared relaxation, and it is correlated with distinctive patterns of hadronic, diamagnetic, and paramagnetic EDMs. These observations motivate targets for future EDM measurements of various systems. In particular, EDM measurements of light nuclei would be very welcome, as the nuclear theory is  under good control and the EDMs are not suppressed by Schiff screening. In addition, a rich experimental program is possible with EDMs and magnetic quadrupole moments of radioactive molecules \cite{Hutzler:2020lmj} to further constrain hadronic CP violation. To make the most of existing and future experiments, hadronic and nuclear theory must be improved. Lattice QCD calculations are underway that connect dimension-six operators, such as the quark chromo-EDM, to nucleon EDMs \cite{Shindler:2021bcx,Kim:2021qae} but still have a long way to go. Improved nuclear structure calculations of, for example, ${}^{199}$Hg and ${}^{225}$Ra EDMs would be very beneficial in connecting observed patterns of EDMs to ultraviolet physics and the nature of the solution to the strong CP problem.

~\\
~\\

{\bf Acknowledgements:}  PD  acknowledges support from the US Department of Energy under Grant No. DE-SC0015655. KF is supported by the US Department of Energy through the Los Alamos National Laboratory and the LANL LDRD Program. Los Alamos National Laboratory is operated by Triad National Security, LLC, for the National Nuclear Security Administration of U.S. Department of Energy (Contract No. 89233218CNA000001).
The work of BL was performed in part at Aspen Center for Physics, which is supported by National Science Foundation grant PHY-1607611. The work of BL was partially supported by a grant from the Simons Foundation.
\bibliography{theta_refs}
\bibliographystyle{utphysmod}

\end{document}